\pdfoutput=1
\RequirePackage{fix-cm}
\documentclass[smallcondensed,final]{svjour3}     
\smartqed  
\usepackage{graphicx}
\usepackage{mathptmx}      
\pdfoutput=1

\usepackage[utf8]{inputenc}
\usepackage[T1]{fontenc}
\usepackage{ae,aecompl}

\usepackage{color}
\definecolor{darkblue}{rgb}{0,0,0.6}
\definecolor{darkred}{rgb}{0.6,0,0}
\definecolor{darkgreen}{rgb}{0,0.6,0}

\usepackage[colorlinks=true,urlcolor=darkblue,citecolor=darkblue,linkcolor=darkred,hyperfootnotes=false]{hyperref}

\usepackage{amsmath,amssymb}

\usepackage{array}

\newcommand{\Vf}{V^{\!f}\!}
\newcommand{\T}{\mathcal T}
\newcommand{\dd}{\!\text{d}}
\newcommand{\ee}{\text{e}}
\newcommand{\cc}{\text{c}}

\renewcommand{\phi}{\varphi}

\newcommand{\yi}{y_{\text{i}}}
\newcommand{\yf}{y_{\text{f}}}
\newcommand{\yfi}{y_{\text{f,i}}}
\newcommand{\hyi}{{\hat y}_{\text{i}}}
\newcommand{\hyf}{{\hat y}_{\text{f}}}
\newcommand{\hyfi}{\hat y_{\text{f,i}}}

\newcommand{\tf}{t_{\text{f}}}

\newcommand{\htf}{\hat t_{\text{f}}}
\newcommand{\FF}{\text{F}}
\newcommand{\tw}{\text{wall}}
\newcommand{\vmoy}{\overline{v}}
\newcommand{\opt}{\text{opt}}
\newcommand{\KPZ}{\text{KPZ}}

\newcommand{\ii}{\text{i}}
\newcommand{\ff}{\text{f}}

\newcommand{\deltay}{\delta\!y}

\usepackage[usenames,dvipsnames]{xcolor}

\journalname{Journal of Statistical Physics}
\begin{document}

\title{Driven interfaces: from flow to creep through model reduction}
\titlerunning{Driven interfaces: from flow to creep through model reduction}

%
%
%


\author{
Elisabeth~Agoritsas$\:^\sharp$
\and
\\{Reinaldo~Garc\'ia-Garc\'ia$\:^\sharp$}
\and
Vivien Lecomte
\and
Lev~Truskinovsky
\and
Damien Vandembroucq
}

\authorrunning{E.~Agoritsas, R.~Garc\'ia-Garc\'ia~\emph{et al.}} 

\institute{
Elisabeth Agoritsas \email{elisabeth.agoritsas@univ-grenoble-alpes.fr} 
\at Universit\'e Grenoble Alpes, LIPHY, F-38000 Grenoble, France
\at CNRS, LIPHY, F-38000 Grenoble, France
\and
Reinaldo Garc\'ia-Garc\'ia  \email{reinaldomeister@gmail.com} \at Laboratoire de Physico-Chimie Th\'eorique-UMR CNRS Gulliver 7083, PSL Research University,
ESPCI, 10 rue de Vauquelin, 75231 Paris cedex 05, France
\and
Vivien Lecomte \email{vivien.lecomte@math.univ-paris-diderot.fr} \at LPMA, UMR 7599, Universit\'e Paris Diderot and Pierre et Marie Curie, Paris, France
\and
Lev Truskinovsky \email{lev.truskinovsky@espci.fr} \and \\
Damien Vandembroucq \email{damien.vandembroucq@espci.fr} 
\at 
PMMH, CNRS UMR 7636, ESPCI, Univ. Pierre et Marie Curie, Univ. Paris Diderot, Paris, France
\and
$^\sharp$ These authors equally contributed to this work.
}

\date{}

\maketitle

\begin{abstract}
\vspace*{-15mm}
The response of spatially extended systems to a force leading their steady state out of equilibrium is strongly affected by the presence of disorder.
We focus on the mean velocity induced by a constant force applied on one-dimensional interfaces. 
In the absence of disorder, the velocity is linear in the force. 
In the presence of disorder, it is widely admitted, as well as experimentally and numerically verified, that the velocity presents a stretched exponential dependence in the force (the so-called `creep law'), which is out of reach of linear response, or more generically of direct perturbative expansions at small force.
In dimension one, there is no exact analytical derivation of such a law, even from a theoretical physical point of view.
We propose an effective model with two degrees of freedom, constructed from the full spatially extended model, that captures many aspects of the creep phenomenology.
It provides a justification of the creep law form of the velocity-force characteristics, in a quasistatic approximation.
It allows, moreover, to capture the non-trivial effects of short-range correlations in the disorder, which govern the low-temperature asymptotics.
It enables us to establish a phase diagram where the creep law manifests itself in the vicinity of the origin in the force{~--~}system-size{~--~}temperature coordinates.
Conjointly, we characterise  the crossover between the creep regime and a linear-response regime that arises due to finite system size.

\keywords{Disordered systems \and Non-equilibrium dynamics \and Creep law  \and Non-linear response \and Kardar--Parisi--Zhang universality class}
\end{abstract}

\setcounter{tocdepth}{2}
\tableofcontents

\bigskip
\section{Introduction}
\label{sec:introduction}

Disorder can radically affect the behaviour of physical phenomena.
An archetypal class of systems is given by extended elastic objects (lines or manifolds) which fluctuate in an heterogeneous medium~\cite{brazovskii_pinning_2004}. Examples range from interfaces in magnetic or ferroic~\cite{kleemann_universal_2007} materials,  vortices in superconductors~\cite{blatter_vortices_1994} to solid membranes in chemical or biological liquids, and fronts in liquid crystals~\cite{takeuchi_universal_2010,takeuchi_evidence_2012}.
In the absence of disorder, the geometry and dynamical properties of such systems are in general resulting from a simple interplay between elastic constraints and thermal noise.
The addition of disorder (impurities, quenched inhomogeneities, space-time noise) can alter the geometry of the interface, by changing it from flat to rough, or by modifying its fractal dimension in scale-invariant systems~\cite{barabasi_fractal_1995}.
For models in the Kardar-Parisi-Zhang class (KPZ) class~\cite{kardar_dynamic_1986} the disorder transforms the diffusive spatial fluctuations of a line into superdiffusive ones (see~\cite{bouchaud_mezard_parisi_1995_PhysRevE52_3656,halpin-healy_kinetic_1995,corwin_kardarparisizhang_2012,agoritsas_static_2013,halpin-healy_kpz_2015} for reviews).

We are interested in the non-equilibrium motion induced by an external drive applied to such systems with quenched disorder.
In the disorder-free situation, Ohm's linear law~\cite{ohm_galvanische_1827} between the observed average velocity and the applied force 
yields a simple linear response.
In the presence of disorder, the situation is more complex; the so-called `creep law' is an example of velocity-force characteristics for which linear response does not hold, even in the very small force limit. It is described by a stretched-exponential velocity-force relation of the form
$  \vmoy(f) 
{\sim}
\ee^{{-f^{-\mu}}}
$
(where we set scaling parameters to 1) depending on the \emph{creep exponent}~$\mu>0$, which is non-analytic at zero force~$f=0$. 
It was derived in the context of dislocations in disordered media
  \cite{ioffe_dynamics_1987,nattermann_interface_1987} and motion of vortex lines \cite{feigelman_theory_1989,feigelman_thermal_1990,nattermann_scaling_1990},
and gave rise to a number of studies, ranging from the initial scaling and renormalization-group (RG) analysis~\cite{kardar_dynamic_1986,huse_huse_1985},
to equilibrium \cite{narayan_threshold_1993} 
and
non-equilibrium functional renormalization group (FRG) studies \cite{chauve_creep_1998,chauve_creep_2000},
and, in the picture of successive activation events, to the study of the distribution of energy barriers~\cite{drossel_scaling_1995}
and of activation times~\cite{vinokur_marchetti_1996_PhysRevLett77_1845,vinokur_glassy_1997}.
Numerical studies confirm its validity for one-dimensional interfaces
~
  \cite{roters_creep_2001,kolton_creep_2005,giamarchi_dynamics_2006,kolton_creep_2009}
~
  (see~\cite{ferrero_2013_ComptesRendusPhys14_641} for a review),
and experimental results for driven domain walls in ultrathin magnetic layers~\cite{lemerle_1998_PhysRevLett80_849} are compatible with a stretched exponential velocity-force relation with a creep exponent $\mu=1/4$.

Several questions yet remain to be clarified.
The first class of questions pertains to the derivation of the law itself: RG analysis in dimension one is known to be non-convergent, and FRG approaches are valid perturbatively in dimension $4-\epsilon$ far from dimension~$1$ ($\epsilon \ll 1$). Different power-counting scaling arguments lead to different values of the creep exponent $\mu$ (as we detail in subsection~\ref{ssec:naivescaling}).
The second class of questions is related to the understanding of the finite-size regime: one indeed expects that for a finite system, the linear response should be valid at very low forces; this rises the question of how to depict the crossover between this linear regime and the creep law.
More generically, we aim at constructing a phase diagram in the three coordinates (force, inverse of system size, temperature) provided by the parameters of interest, which would depict criticality around its origin and specify the characteristic scales of the creep regime.

In this article, we construct an effective model describing the motion of the driven interface at fixed lengthscale. It allows us to recover the creep law and to extend the description of the dynamics from low forces to larger forces, and to characterise the crossover between creep and linear response in finite systems in the very small force regime.
Previous approaches
have dealt with `zero-dimensional' toy models where a particle with one degree of freedom moves in a one-dimensional random potential~\cite{vinokur_glassy_1997,le_doussal_creep_1995,scheidl_mobility_1995}.
However, the distribution of such a random potential that would summarise the effect of the disorder experienced by the full segment of the interface would prove very delicate to describe in our system of interest. Indeed, an important issue of driven systems with one degree of freedom is that very large barriers always block the motion, making the extreme statistics of the disorder play an essential role. In contrast, the cost of the elastic deformations allowed in higher dimensions allows to counterbalance large inhomogeneities in the environment:   deep wells in the disorder potential cannot pin the interface beyond the (unbounded) energetic cost of the elastic deformations that they would impose as the interface moves ---~thus rendering the dynamics of the system less sensitive to the extremes of the disorder distribution.
The effective model that we propose has two degrees of freedom which allow to capture in a minimalist way such competition between elasticity and disorder.

\begin{figure}[t]
  \centering
  \includegraphics[width=.95\columnwidth]{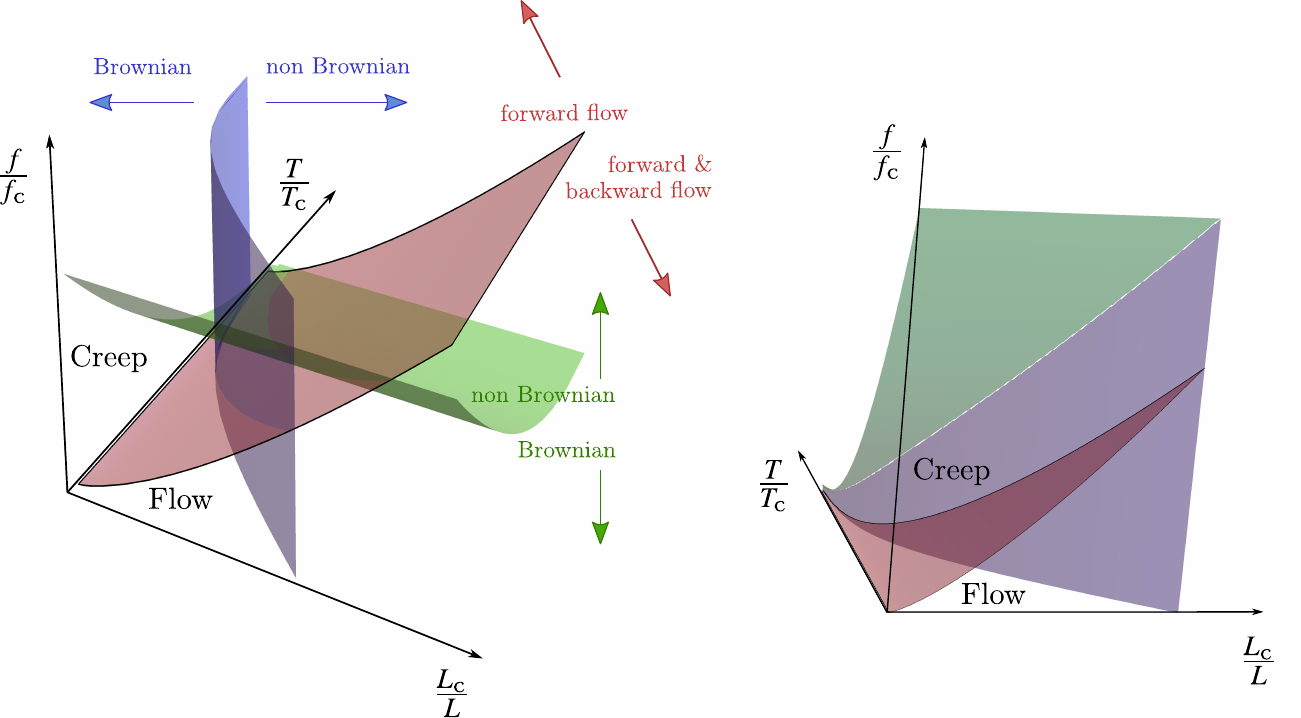}
  \caption{
(\textbf{Left})
Phase diagram of the regimes of velocity-force dependence. Coordinates are the inverse reduced system size $L_\cc/L$, the reduced temperature $T/T_\cc$ and the reduced force $f/f_\cc$. The creep law holds in a region in the vicinity of the origin.
The manifold of equation $f/f_\cc = (L_\cc/L)^{4/3}$ (in \textbf{red}) separates the creep regime from the flow regime where finite system size hinders the scale invariance leading to the creep law, and induces instead a linear response velocity-force characteristic. Above this manifold, the motion is dominated by forward motion in the direction of the drive while, below, both the forward and backward motions play a comparable role.
The manifold in \textbf{blue} (equation given by~\eqref{eq:bluemanifold}) separates the regime of system size where a Brownian scaling of free-energy ensures the validity of the creep law form (for large enough $L$) from a regime not govern by the creep law. 
It provides an upper bound to the validity of the scaling analysis.
The manifold in \textbf{green} (equation given by~\eqref{eq:green-manifold}) separates the regimes of force where, similarly, a Brownian scaling of free-energy holds or not, when, respectively, the typical interface length governing the motion is larger or smaller than the Larkin length. 
It provides an upper bound to the validity of the scaling analysis.
The results are derived in Sec.~\ref{ssec:phasediag} and~\ref{ssec:tempescaling} together with the expressions of the characteristic length~$L_\cc$, the characteristic temperature $T_\cc$ and the characteristic force~$f_\cc$ in terms of the model parameters (see also Table~\ref{tab:notations}).
Most of the recent advances on exact characterisations of the distribution of free-energy in KPZ (reviewed \emph{e.g.} in~\cite{corwin_kardarparisizhang_2012,halpin-healy_kpz_2015}) are restricted to uncorrelated disorder, \emph{i.e.}~lie in the $T\gg T_\cc$ region, far from the creep regime.
(\textbf{Right}) Zoom of the resulting domains in the vicinity of the origin.
}
  \label{fig:phase-diagram}
\end{figure}

We now summarise our findings, before providing their derivation in the next sections.
We establish a phase diagram in the force~--~system-size~--~temperature coordinates
(see Fig.~\ref{fig:phase-diagram})
that describes the regimes where the velocity-force dependence is of the creep type or of the flow type.
The creep law itself appears in the critical region of the diagram, in the low-force, low-temperature and large-system size regime.
Physically, the creep regime holds in regions where a specific form of scale invariance holds (relevant observables scaling simply with all parameters), while the flow regime appears whenever the system-size is too small for such scaling to hold.
Our results complement previous numerical and phenomenological studies on the scalings of the driven interface~\cite{kolton_creep_2009,ferrero_2013_ComptesRendusPhys14_641,ferrero_2016_ArXiv-1604.03726}: we are able to analyse the role of finite system size (complementing~\cite{kim_interdimensional_2009}), but we cannot probe the depinning regime that was examined in  those studies,
{which is out of reach of our effective model reduction}.
The novelty of our approach lies also in the methodology that we propose, which provides a
well-defined procedure, solving in particular a power-counting dilemma for the scaling arguments:
we device an effective model for the driven interface \emph{at fixed scale} and propose
 saddle-point point argument at small forces which justifies why only one out of two possible power-counting arguments (either on the Hamiltonian or on the free energy) yields the correct creep exponent.
We
detail how a mean first passage time (MFPT) description allows to handle non-equilibrium issues within an equilibrium settings.
Then, from the combination of those tools, we are able  to extend the creep law and we describe the crossover from the finite-system size very small velocity regime to the creep law.
Most importantly, we take into account the role of finite disorder correlations at short-range~\cite{agoritsas_static_2013,nattermann_interface_1988,agoritsas_temperature-induced_2010,agoritsas_disordered_2012,agoritsas_temperature-dependence_2013}
which are essential to understand the low-temperature asymptotics, and in particular to identify the characteristic energy and characteristic force of the creep law.

The article is organised as follows:
we precise the model and known results in Sec.~\ref{sec:model}.
We review the particle toy-model (with one degree of freedom) in Sec.~\ref{sec:particle} as it serves as a basis to our analysis.
In Sec.~\ref{sec:creep}, we describe the effective model with two degrees of freedom and explain how it provides a useful framework to understand the creep law.
In Sec.~\ref{sec:finite-size}, we use this description to analyse the crossover from creep to linear response at small forces in finite systems.
We finally discuss our results in Sec.~\ref{sec:discussion} and present perspectives 
in Sec.~\ref{sec:conclusion}.
Our exposition is self-contained; the reader familiar with the subject can read in Sec.~\ref{ssec:tiltedDPf} the definition of the tilted KPZ problem and directly jump to Sec.~\ref{sec:creep} for the analysis the effective model derived from it.
Table~\ref{tab:notations} summarises the notations.

\begin{table}[h]
  \centering
  \begin{tabular}{|l|lllc|}
\hline
&  \textbf{Variable}  & \textbf{Signification} & \textbf{Expression} & \textbf{Eq./Fig./§} 
\\
\hline
\textbf{Coordinates}   
&  $y$                & Transverse coordinate   &                     & 
\\
&  $t$                & Longitudinal coordinate &                     & 
\\
&  $\tau$             & Physical time           &                     &  Fig.~\ref{fig:interface}
\\
&  $\tf$              & Interface segment length&                     & 
\\
&  $\yi,\yf$          & Starting and arrival points     &             & 
\\
\hline
\textbf{Model}   
&  $c$                & Elastic constant        &                     & \eqref{eq:Langevin_yttau_forcef}
\\
\textbf{parameters}   
&  $D$                & Disorder strength       &                     & \eqref{eq:VVxi}
\\
&  $T$                & Temperature             &                     & §~\ref{ssec:dyn_int}
\\
&  $\xi$              & Disorder correlation length&                  & \eqref{eq:VVxi}
\\
&  $f$                & Driving force           &                     & \eqref{eq:Langevin_yttau_forcef}
\\
&  $\gamma$           & Friction coefficient    &                     & \eqref{eq:Langevin_yttau_forcef}
\\
&  $L$                & Total interface length  &                     & §~\ref{ssec:fsize}
\\
\hline
\textbf{{Thermodynamic}}   
&  $\vphantom{\Big|}W_V^f$            & Partition function      &                      & \eqref{eq:defWVf}
\\
\textbf{{quantities}}   
&  $\vphantom{\Big|}F_V^f$            & Free energy             &                      & \eqref{eq:defFfvinit}
\\
&  $\vphantom{\Big|}\bar F_V^f$       & Disorder free energy    &                      & \eqref{eq:STSwithf_forMFPT}
\\
\hline
\textbf{{Observables}}
&	$\vphantom{\Big|} \bar{v}(f)$		& {Steady-state velocity}	&	& \eqref{eq:creep_1d_TM}, \eqref{eq:creep_1d_finiteL_sh}
\\
&	$\vphantom{\Big|} \bar\tau_1(f)$	& {Mean First Passage Time}	&	& \eqref{eq:restau1_non-ave}, (\ref{eq:vftimeintegral}-\ref{eq:finaltau1})
\\
\hline
\textbf{Effective}   
&  $\tilde \gamma$    & Effective friction      &   $\vphantom{\Big|}\tfrac 12 \tf\,\gamma$    & \eqref{eq:Ttildegammatilde}
\\
 \textbf{parameters}
 &  $\tilde T$        & Effective temperature   &   $\vphantom{\Big|}T$                                        & \eqref{eq:Ttildegammatilde} 
\\
 \hline
 \textbf{Characteristic}   
 &  $T_\cc$            & Characteristic temperature &     $\vphantom{\Big|}(\xi c D)^{1/3}$                      & \eqref{eq:lowhighTregimes}
 \\
 \textbf{parameters}   
 &  $L_\cc$            & Larkin length (at low~$T$)&   $\big[{c^2\xi^5}/{D}\big]^{1/3}$          & \eqref{eq:defLclowT} 
 \\
 &  $f_\cc$            & Characteristic force      &     $\big[{D^2}/({c \xi^7)}\big]^{1/3}$     & \eqref{eq:UcfclowT}
\\
 & $U_\cc$             & Characteristic barrier    &      $\vphantom{\Big|}(g\,T_\cc/T)^{3/4}\, T_\cc$      & \eqref{eq:UcfclowT},\eqref{eq:UcfcfiniteT}
 \\
 & $\vphantom{\big|}F_\cc$ & Critical depinning force       &                                            & Fig.~\ref{fig:interface}
 \\
 \hline
 \textbf{Finite-$\xi$}   
 &  $\vphantom{\Big|}\tilde D$   & Disorder free-energy strength & $\tilde D = g\,{cD}/{T}$   & (\ref{eq:rescaling_free-energy_FbarfV}-\ref{eq:lowhighTregimes}),\eqref{eq:Dtildevsfudgingg}
 \\
 \textbf{parameters}   
 &  $g$                & Fudging parameter         &    solution of~\eqref{eq:fudging_eq}      & \eqref{eq:fudging_eq} 
 \\
 &  $L_\opt$            & Optimal length            &    $(c \tilde D )^{1/4} f^{-3/4}$           & \eqref{eq:Lopt_topt_f34}
 \\
 &  $\mathcal L_\cc$   & $T$-dependent Larkin length     &   ${T^5}/({cD^2}g^{5})$                    & \eqref{eq:TdepLarking} 
 \\
 &  $f_L$              & Max. force until $L_\opt(f)=L$& $  (c \tilde D )^{1/3}\,L^{-4/3} $       & \eqref{eq:eqfL}
 \\
 &  $f_{\mathcal L}$     & Max. force until $L_\opt(f)=\mathcal L_\cc$& $ \tilde D^7/(c^5 D^4) $       & \eqref{eq:eq-for-fL}
 \\
 &  $\bar\gamma$       & Fudging exponent          &    $\in\{\tfrac 32,6\}$                    & \eqref{eq:fudging_eq} 
 \\
\hline
  \end{tabular}
  \caption{Table of parameters and notations.}
  \label{tab:notations}
\end{table}

\section{Model and questions}
\label{sec:model}

We present in this section the model of one-dimensional (1D) interface that we consider, recalling its known phenomenology 
in Sec.~\ref{ssec:dyn_int}.
We describe the correspondence, in the non-driven case, between the equilibrium distribution of the position of the interface and the directed polymer 
in Sec.~\ref{ssec:DP}, motivated by understanding the fluctuations of the interface \emph{at fixed length}~{$\tf$} ---~a crucial step for our scaling analysis and that has a natural formulation in the directed polymer language.
We then construct
in Sec.~\ref{ssec:tiltedDPf}
a variation of the equilibrium problem in which the interface is subjected to a tilted random potential but also to boundary conditions forbidding the development of a non-zero velocity state, and that we will use as a starting point in our approach in the following sections.
Finally, to motivate our study, we compare 
in Sec.~\ref{ssec:naivescaling}
different power-counting arguments presented in the literature either at the Hamiltonian or at the free-energy level, which do not lead to the same result and call for a detailed analysis.

\subsection{Dynamics}
\label{ssec:dyn_int}

\begin{figure}
\centering  \includegraphics[width=0.92\textwidth]{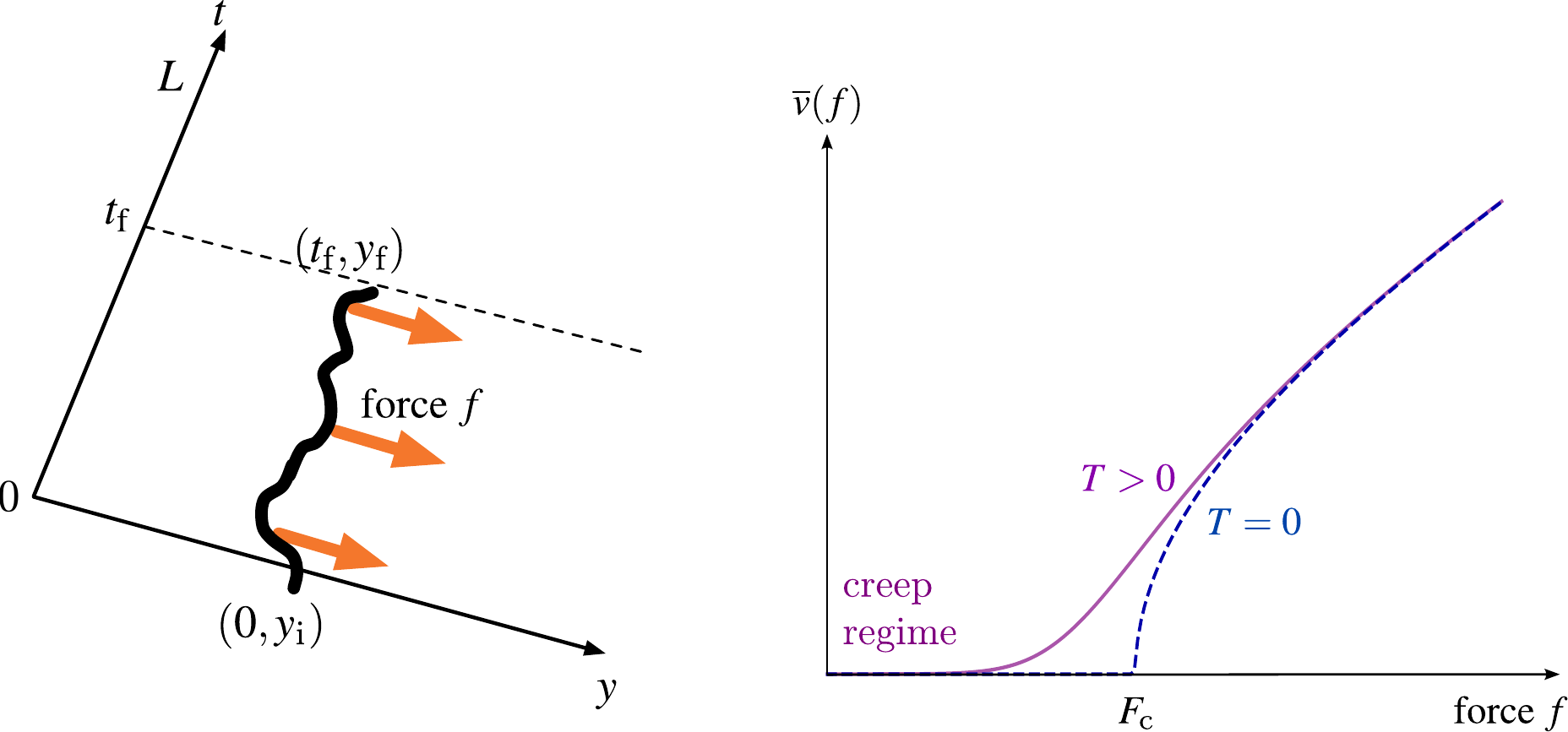}
\caption{ (\textbf{Left})
Trajectory of an interface $y(t,\tau)$ of length $\tf$ at fixed physical time $\tau$. The longitudinal spatial coordinate $t$ of the interface can be understood as ``growth time'' coordinate of a directed polymer. The interface, described by the evolution equation~\eqref{eq:Langevin_yttau_forcef}, is subjected to a force field $f$
which induces a mean velocity $\vmoy(f)$ of the average interface position. The total system length along direction $t$ is denoted $L$.
(\textbf{Right})~Schematic representation  of the velocity-force characteristic $\overline{v}(f)$. At zero temperature (dashed blue) the interface remains pinned until the depinning transition at a force $F_\cc$. At non-zero temperature (purple continuous line), the initial regime at $f\ll F_\cc $ is very slow and obeys the creep law~\eqref{eq:thecreeplaw-mu}. At $f=F_\cc$ the $T=0$ characteristic is rounded by temperature. At large force $f\gg F_\cc$ the temperature and disorder play no role and the velocity becomes linear in $f$.
}
\label{fig:interface}
\end{figure}

We denote by $\tau$ the (physical) time of the interface and introduce as described in Fig.~\ref{fig:interface} a $\tau$-dependent position $y(t,\tau)$
of the interface, of longitudinal coordinate~$t$.  Its evolution is described by the overdamped Langevin equation 
\begin{align}
  \gamma \partial_\tau y(t,\tau) &= 
  c\partial_t^2y -\partial_yV(t,y(t,\tau))+f+\eta(t,\tau)
  \label{eq:Langevin_yttau_forcef}
\end{align}
where $c$ is the elastic constant, $\gamma$ the friction coefficient and $f$~the driving force.
Thermal fluctuations at temperature~$T$ are described by the centred white noise $\eta(t,\tau)$ of correlations $\langle\eta(t',\tau')\eta(t,\tau)\rangle=2\gamma T\delta(t'-t)\delta(\tau'-\tau)$. 
We have set Boltzmann's constant to $k_{\text{B}}=1$.
The disorder $V(t,y)$ has a Gaussian distribution of zero mean and correlations fully described by its two-point function
\begin{equation}
  \overline{V(t',y')V(t,y)} \ = \ D \delta(t'-t) R_\xi(y'-y)
  \label{eq:VVxi}
\end{equation} 
Longitudinal correlations in the direction $t$ are absent while transverse correlations are described by a function $R_\xi(y)$ scaling as $R_\xi(y)=\frac 1\xi \hat R_1(y/\xi)$ and normalised as $\int_{\mathbb R} \dd y\,R_\xi(y)=1$. More specifically, the correlator $R_\xi(y)$ is a ``smooth delta'' describing short-range correlations at scale~$\xi$, that tends to a Dirac delta as $\xi$ goes to zero. The strength of disorder is described by the parameter $D$. Such type of quenched disorder belongs to the `random bond' class.

The driving force $f$ induces a motion of the interface, characterised by its mean velocity $\vmoy(f)$.
The linear response fails for an infinite interface even in the small-force regime, and instead of a velocity proportional to $f$ one observes the `creep law'
\begin{equation}
  \vmoy(f) 
\underset{f\to 0}{\sim}
\ee^{\text{
$
\displaystyle{-\frac {U_\cc}T \Big(\frac{f_\cc}f\Big)^\mu}
$
}}
\qquad\text{with}\qquad
 \mu=\frac 14
\label{eq:thecreeplaw-mu}
\end{equation}
This is the stretched exponential behaviour, already mentioned in the introduction, of creep exponent $\mu$, characteristic energy scale $U_\cc$ and characteristic force~$f_\cc$.
It is valid in the regime of low forces compared to the depinning critical force $F_\cc$ and of low temperatures compared to the effective barrier $U_\cc\;(f_\cc/f)^{1/4}$ (see Fig.~\ref{fig:interface}, Right).
The characteristic parameters $f_\cc$ and $U_\cc$ are usually {identified} 
numerically or by a fitting procedure; in this article, we {derive} 
an expression of $f_\cc$ and $U_\cc$ in terms of the model parameters $c$, $D$, $\xi$ valid in the low-temperature limit of the 1D interface (see~Sec.~\ref{ssec:scalingandthecreeplaw}), {including a non-trivial temperature dependence.}
The behaviour~\eqref{eq:thecreeplaw-mu} has been verified experimentally for interfaces in magnetic materials on several decades of velocities~\cite{lemerle_1998_PhysRevLett80_849} and tested numerically with success~\cite{kolton_creep_2005,giamarchi_dynamics_2006,kolton_creep_2009,ferrero_2013_ComptesRendusPhys14_641}.
It was originally predicted in other dimensionalities in the description of the motion of vortices in random materials {(modelling for instance vortices in type-II superconductors)}, using either scaling~\cite{ioffe_dynamics_1987,nattermann_scaling_1990} 
or perturbative FRG arguments in an expansion around spatial dimension 4,
first used within an equilibrium frame~\cite{narayan_threshold_1993} and then extended out of equilibrium~\cite{chauve_creep_1998}.%
%

\subsection{Zero driving force: correspondence with the directed polymer}
\label{ssec:DP}

We first consider the equilibrium case at zero driving force~$f=0$, which has been extensively studied (see~\cite{halpin-healy_kinetic_1995,corwin_kardarparisizhang_2012,agoritsas_static_2013} for reviews). The Langevin equation becomes
\begin{align}
  \gamma \partial_\tau y(t,\tau) &= 
  c\partial_t^2y -\partial_yV(t,y(t,\tau))+\eta(t,\tau)
  \label{eq:Langevin_yttau}
\end{align}
At fixed disorder $V$, the system eventually reaches an equilibrium steady state in the long-time limit: the
non-normalised weight of a segment of interface $y(t)_{0\leq t\leq \tf}$ of length~$\tf$ is $\ee^{-\frac 1T \mathcal H_V[y]}$, where the Hamiltonian $\mathcal H_V[y]$ 
reads
\begin{equation}
\mathcal H_V[y] = \int
 dt\:  \Big[ \frac c2 (\partial_ty)^2+V(t,y(t))\Big]
\label{eq:defHVf0}
\end{equation}
where the boundaries of the integral are determined by the domain of definition of~$y(t)$.
The Boltzmann equilibrium form of this steady state can directly be read from the equation of evolution~\eqref{eq:Langevin_yttau} that one can rewrite as
\begin{align}
  \gamma \partial_\tau y(t,\tau) &= 
  -\frac{\delta \mathcal H_V[y(\cdot,\tau)]}{\delta y(t,\tau)}  + \eta(t,\tau)
  \label{eq:Langevin_yttau_dHdy}
\end{align}
allowing to recognise an overdamped Langevin dynamics of force term deriving from the Hamiltonian~\eqref{eq:defHVf0}.
One can explicitly check from the functional Fokker-Planck equation associated to~\eqref{eq:Langevin_yttau}:
\begin{equation}
 \gamma
 \partial_\tau \mathbb P [y(\cdot),\tau]= \int d t
 \frac{\delta}{\delta y(t)}
 \Big\{
  \frac{\delta \mathcal H_V[y]}{\delta y(t)} \mathbb P [y,\tau]  + T \frac{\delta \mathbb P [y,\tau]}{\delta y(t)}
 \Big\}
\end{equation}
that the distribution
\begin{equation}
\mathbb P_\text{eq} [y]\propto \ee^{-\frac 1T \mathcal H_V[y]}
\end{equation}
is a zero-probability-current steady-state solution (hence an equilibrium one)
\begin{equation}
 0= \int d t
 \frac{\delta}{\delta y(t)}
 \Big\{
  \underbrace{\frac{\delta \mathcal H_V[y]}{\delta y(t)} \mathbb P_\text{eq} [y]  + T \frac{\delta \mathbb P_\text{eq} [y]}{\delta y(t)}}_{=0}
 \Big\}
\label{eq:funcFPzerof}
\end{equation}
Spatial boundary conditions then determine how the Boltzmann weight comes into play when defining probability distributions; a well-understood situation is that of the continuous directed polymer, where the interface $y(t)$ is attached at its extremities in $t=0$ and $t=\tf$ (adopting so-called \emph{point-to-point} configurations, see Fig.~\ref{fig:interface}, left). The equilibrium weight at temperature~$T$ of realisations of the interface starting from $(0,\yi)$ and arriving in $(\tf,\yf)$ in a random potential $V(t,y)$ is the weight $W_V(\tf,\yf|0,\yi)$ defined by the path integral:
\begin{equation}
  W_V(\tf,\yf|0,\yi) = \int_{y(0)=\yi}^{y(\tf)=\yf}\mathcal Dy(t)\: 
  \exp\Big\{-\frac 1T
    \int_{0}^{\tf} dt\:  \Big[ \frac c2 (\partial_ty)^2+V(t,y(t))\Big]
  \Big\}
  \label{eq:defWV}
\end{equation}
For instance, at fixed length~$\tf$ and fixed initial position~$\yi$, the probability density of interfaces arriving in $\yf$ is $W_V(\tf,\yf|0,\yi)/\int d\tilde y\;W_V(\tf,\tilde y|0,\yi)$.
The statistical properties of the `partition function'~\eqref{eq:defWV} have been the subject of extensive studies, as the continuous directed polymer belongs to the Kardar-Parisi-Zhang (KPZ)~\cite{kardar_dynamic_1986} universality class of models (see~\cite{bouchaud_mezard_parisi_1995_PhysRevE52_3656,halpin-healy_kinetic_1995,corwin_kardarparisizhang_2012,agoritsas_static_2013,halpin-healy_kpz_2015} for reviews). A noticeable fact is that the directed polymer free energy
\begin{equation}
  F_V(\tf,\yf|0,\yi) = -T\log W_V(\tf,\yf|0,\yi)
  \label{eq:defFvinit}
\end{equation}
verifies the KPZ equation~\cite{huse_huse_1985} with sharp-wedge initial condition: denoting for short $F_V(t,y)=F_V(t,y|0,\yi)$ one has
\begin{equation}
  \partial_t F_V(t,y)= \frac{T}{2c} \partial_y^2 F_V(t,y) -\frac{1}{2c}\big[\partial_yF_V(t,y)\big]^2 + V(t,y)
 \label{eq:theKPZeq}
\end{equation}
We will present and/or derive its useful symmetries for our study when needed (see Appendix~\ref{sec:app_STS} and also Ref.~\cite{canet_nonperturbative_2011} for a systematic study of KPZ symmetries).
Note that a proper mathematical definition of~\eqref{eq:defWV} as an expectation over Brownian bridges~\cite{amir_probability_2011} and the passage to the KPZ equation~\eqref{eq:theKPZeq} through the application of It{\=o}'s lemma requires the appropriate removal of  $(\xi\to 0)$-diverging constants $\propto R_\xi(0)$ (see~\cite{quastel_introduction_2011} for a pedagogical introduction).

\subsection{The tilted directed polymer: equilibrium at non-zero $f$}
\label{ssec:tiltedDPf}

The original dynamics of the driven interface~\eqref{eq:Langevin_yttau_forcef} at non-zero driving force $f$ can also be rewritten in a form similar to~\eqref{eq:Langevin_yttau_dHdy} 
\begin{align}
  \gamma \partial_\tau y(t,\tau) &=  
  -\frac{\delta \mathcal H_V^f[y(\cdot,\tau)]}{\delta y(t,\tau)}  + \eta(t,\tau)
  \label{eq:Langevin_yttau_forcef_withH}
\end{align}
as an overdamped Langevin equation with forces deriving from a \emph{tilted} Hamiltonian
\begin{equation}
  \mathcal H_V^f[y]\equiv 
  \int
  dt\:  \Big[ \frac c2 (\partial_ty)^2+V(t,y(t))-fy(t)\Big]
\label{eq:defHVfnon0}
\end{equation}
One can still write a Boltzmann equilibrium distribution
$
\mathbb P_\text{eq} [y]\propto \ee^{-\frac 1T \mathcal H_V^f[y]}
$
which is a zero-probability-current solution to the steady-state functional Fokker-Planck equation, similarly to what we observed in~\eqref{eq:funcFPzerof} (with $\mathcal H_V\mapsto \mathcal H_V^f$).
However, it describes the steady state of the system only  in situations where the dynamics is \emph{reversible}.
In the presence of the drive~$f$, such an equilibrium can only be reached with appropriate spatial boundary conditions, for instance for a (half-)bounded system with one wall ($-\infty <y(t,\tau)<Y_\tw$ for $f>0$, $Y_\tw<y(t,\tau)<\infty$ for $f<0$) or two walls ($Y_\tw^- <y(t,\tau)<Y_\tw^+$).

Such boundary conditions block the motion of the interface and make average velocity $\vmoy(f)$ equal to zero~; consequently, in the large time limit, they cannot depict the steady state of the non-equilibrium driven interface.
However, as we will argue in Sec.~\ref{sec:creep}, \emph{the distribution of point-to-point interfaces in this $f\neq 0$ equilibrium settings still provides an effective model for the non-equilibrium quasistatic motion of the driven interface}. To construct this effective model, the central quantity that we will use is the
weight of a trajectory starting from $(0,\yi)$ and arriving in $(\tf,\yf)$ in a tilted random potential $\Vf(t,y)\equiv V(t,y)-fy$:
\begin{equation}
  W_V^f(\tf,\yf) = \int_{y(0)=\yi}^{y(\tf)=\yf}\mathcal Dy(t)\: 
  \exp\Big\{-\frac 1T
    \int_{0}^{\tf} dt\:  \big[ \frac c2 (\partial_ty)^2+V(t,y(t))-fy(t)\big
    ]
  \Big\}
  \label{eq:defWVf}
\end{equation}
The path integral is performed over the interface configurations respecting the equilibrium boundary conditions with one or two walls.
Similarly to~\eqref{eq:defFvinit} we define a tilted free energy 
\begin{equation}
  F^f_V(\tf,\yf|0,\yi) = -T\log W^f_V(\tf,\yf|0,\yi)
  \label{eq:defFfvinit}
\end{equation}
in which one can precisely identify the contributions scaling differently from each other, as we detail in~\ref{ssec:MFPTeff}.

\subsection{Issues occurring when scaling the Hamiltonian}
\label{ssec:naivescaling}

Before describing the symmetries of the above introduced models, we discuss the scaling arguments based on the Hamiltonian~\eqref{eq:defHVfnon0} that are usually put forward to derive the creep law in an formal way, and how they can present an inconsistency.
In standard heuristic arguments on the scaling properties of the Hamiltonian (see~\cite{giamarchi_dynamics_2006} for a review), it is assumed that the fluctuations of a segment of length $L$ of the interface scale according to $y(t)\sim L^\zeta$ with~$\zeta$ the \emph{roughness exponent} of the interface.
Accordingly, the elastic, disorder and driving contributions to the tilted Hamiltonian~\eqref{eq:defHVfnon0} scale respectively as
\begin{align}
  \mathcal H_\text{el}[y] =
  \int_{0}^{L} dt\:  \frac c2 (\partial_ty)^2 
  & \ \sim \ L^{2\zeta-1} f\,^0
\label{eq:Hel-standardscalingL}
\\
  \mathcal H_\text{dis}[y] =
  \int_{0}^{L} dt\:  V(t,y(t))
   &\ \sim \ L^{\frac{1-\zeta}{2}} f\,^0 
\label{eq:Hdis-standardscalingL}
\\
  \mathcal H_V^f[y] = -
  \int_{0}^{L} dt\: fy(t)
   &\ \sim \ L^{1+\zeta} f
\label{eq:Hforce-standardscalingL}
\end{align}
where for the disorder contribution~\eqref{eq:Hdis-standardscalingL} one uses ${V(L\hat t, L^\zeta\hat y) \stackrel{\text{(d)}}{=} L^{-(1+\zeta)/2}V(\hat t, \hat y)}$ (the symbol $\stackrel{\text{(d)}}{=}$ meaning that the scaling holds in distribution).
Matching the elastic and driving contributions~\eqref{eq:Hel-standardscalingL} and~\eqref{eq:Hforce-standardscalingL} gives the scaling of the optimal interface length $L_\opt$ displaced at a force~$f$
\begin{equation}
  L^{2\zeta-1} f^0 \sim L^{1+\zeta} f \quad \Rightarrow \quad L=L_\opt \sim f^{-\frac{1}{2-\zeta}}
\end{equation}
In this argument, it is then asserted that the average velocity scales as the inverse of the Arrhenius time to cross an energetic barrier, itself scaling as one of the contributions~(\ref{eq:Hel-standardscalingL}-\ref{eq:Hforce-standardscalingL}) to the Hamiltonian. Using either $\mathcal H_\text{el}$ or $\mathcal H_\text{force}$ yields by definition of $L_\opt$ the same result
\begin{equation}
  \vmoy(f)\ \sim \ \ee^{-\frac 1T f^{-\mu}}
\quad\text{with}\quad
  \mu = \frac{-1+2\zeta}{2-\zeta}
\end{equation}
This expression of the creep exponent $\mu$ matches for $d=1$ the generic result known in dimension $d$: $ \mu = \frac{2-d+2\zeta}{2-\zeta}$ (see~\cite{giamarchi_dynamics_2006} for a review).
Substituting the KPZ roughness exponent $\zeta_\KPZ=2/3$, one finally obtains the expected creep exponent~$\mu=1/4$.

Nevertheless, this power-counting procedure lacks a proper justification, and in fact presents an inconsistency.
The roughness exponent that it would imply is incorrect; indeed, matching the elastic and disorder contributions~\eqref{eq:Hel-standardscalingL} and~\eqref{eq:Hdis-standardscalingL} yields $2\zeta-1=(1-\zeta)/2$ hence $\zeta=\zeta_\FF\equiv\frac 35$. This is the so-called Flory exponent of the Hamiltonian, different from the exact value~$\zeta_\KPZ$ which characterises the geometrical fluctuation of the 1D interface at large scales.
Besides, even if one decides to impose the value $\zeta=\zeta_\KPZ$ and that one tries to find $L_\opt$ by matching the disorder contribution~\eqref{eq:Hdis-standardscalingL} (instead of the elastic one) to the driving contributions~\eqref{eq:Hforce-standardscalingL}, one gets
\begin{equation}
  L^{\frac{1-\zeta}{2}} f^0 \sim L^{1+\zeta} f 
  \quad \Rightarrow \quad L=L_\opt {\sim} f^{-\frac{2}{1+3\zeta}}
\end{equation}
which would yield incorrectly $L_\opt\sim f^{-5/7}$ and $\mu=\frac 17$.

At $f=0$, for the determination of the roughness exponent $\zeta$, the origin of that problem has been elucidated in Ref.~\cite{agoritsas_static_2013}: in fact one cannot assume that $y(t)$ typically scales with the same exponent at all lengthscales, as was done in~(\ref{eq:Hel-standardscalingL}-\ref{eq:Hforce-standardscalingL}).
This is seen for instance by the result that the roughness function $B(\tf)$, which describes the variance of the endpoint fluctuations for an interface of length $\tf$ scales with different roughness exponents in the $\tf\to\infty$ and $\tf\to 0$ regimes. 
At our knowledge, at the moment, there is no direct and exact way of adapting the rescaling of the Hamiltonian in order to understand these scalings and/or to obtain the correct value of $\zeta$.
It was in fact shown in Ref.~\cite{agoritsas_static_2013} that a different scaling analysis, based on the scaling of the free energy \emph{at fixed lengthscale} is required to derive the value of $\zeta_\KPZ=2/3$.

On the other hand, a different approach consists in performing this power-counting argument not at the Hamiltonian but at the free-energy level~\cite{nattermann_scaling_1990} (see~\cite{gorokhov_diffusion_1998} for a review) and this time it yields the correct value for the roughness exponent $\zeta$ and for the creep one. Such argument however relies on several hypotheses, namely \emph{(i)} that the cost in free energy due to the driving force is linear in $f$, which is not obvious since linear response does not yield the correct velocity, \emph{(ii)} that the power-counting analysis does yield the correct scales for the low-$f$ regime and \emph{(iii)} that the low-temperature asymptotics is well-defined ---~which is non-trivial because already in the $f=0$ case such asymptotics crucially depends on having~$\xi>0$~\cite{agoritsas_static_2013,nattermann_interface_1988,agoritsas_temperature-induced_2010,agoritsas_disordered_2012,agoritsas_temperature-dependence_2013}.
We do not detail this argument here, because the construction we propose in this article will provide a justification to the above hypotheses.

The standard heuristic procedures described above present an arbitrariness, that we aim at clarifying. To proceed, we first consider in the next section the more simple zero-dimensional system of a particle driven in a one-dimensional random potential, and then define in Sec.~\ref{sec:creep} an effective description of the interface at a fixed scale, consisting in two degrees of freedom instead of a continuum, which still allows to derive the creep law and an extension of it that we present in Sec.~\ref{sec:finite-size}.

%
\section{A warming up: the particle in a 1D random potential}
\label{sec:particle}

\begin{figure}%
\centering  \includegraphics[width=0.75\textwidth]{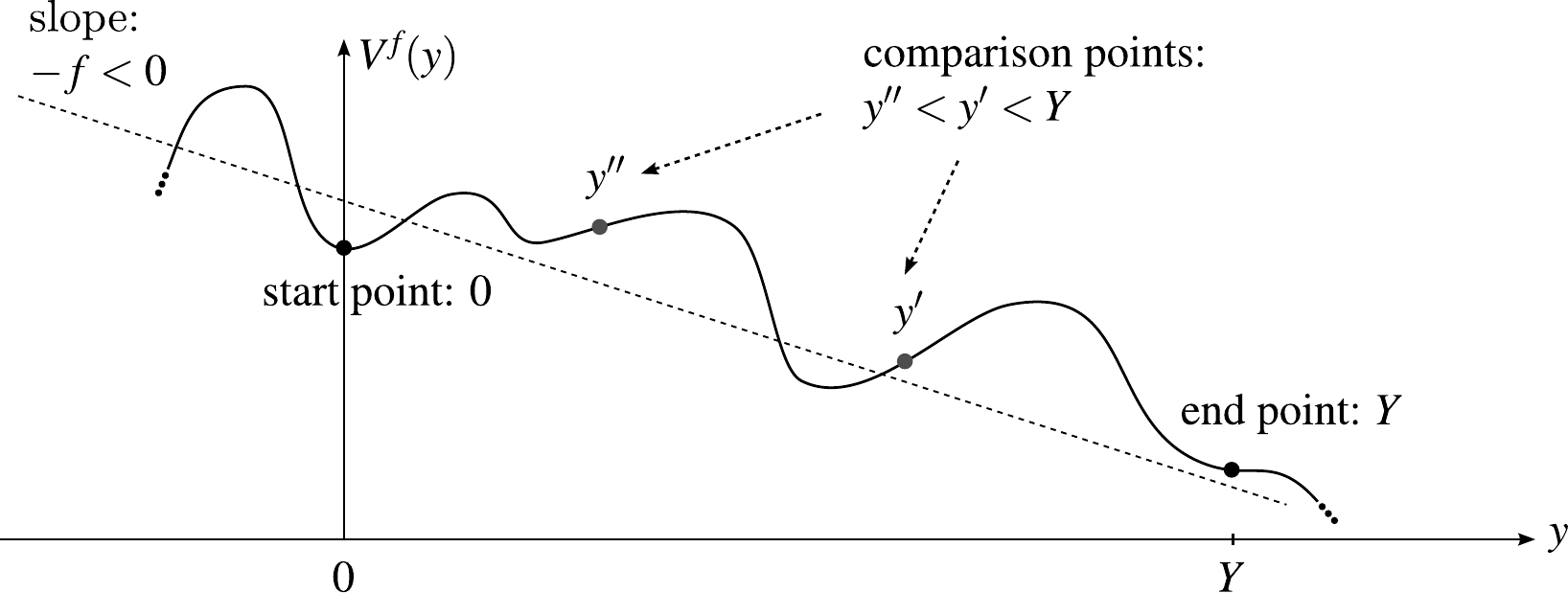}
\caption{
  The tilted potential $\Vf(y)$ in which a particle of position $y(\tau)$ diffuses according to the Langevin equation~\eqref{eq:langevin0D}.
  Its mean velocity $\vmoy(f)$ at large times can be deduced from the exact expression~\eqref{eq:FPT_Boltzmann} of the mean first passage time
    between points $0$ and $Y$, which averages the
    Arrhenius transition times ${Z_V^f(y'')}/{Z_V^f(y')}$ between two comparison points $y''<y'$ such that ${y''<y'<Y}$ and $y'>0$.
  }
\label{fig:landscape_0D}
\end{figure}

The case of a single particle of position $y(\tau)\in\mathbb R$ in an arbitrary potential $V(y)$ and subjected to a driving force~$f>0$ (see Fig.~\ref{fig:landscape_0D}) has been be solved~\cite{le_doussal_creep_1995,scheidl_mobility_1995,gardiner_handbook_1994,risken_fokker-planck_1996} owing to its one-dimensional geometry.
The non-equilibrium steady state of the Fokker-Planck equation associated to the Langevin equation
\begin{equation}
  \gamma\partial_\tau y(\tau) = - \partial_y V(y(\tau))+f+\eta(\tau)
  \label{eq:langevin0D}
\end{equation}
can be obtained exactly~\cite{risken_fokker-planck_1996}. Le Doussal and Vinokur~\cite{le_doussal_creep_1995} and Scheidl~\cite{scheidl_mobility_1995} have used this knowledge of the steady state to determine the mean velocity
$\vmoy(f)=\overline{\partial_\tau \langle y(\tau)\rangle}$ in the steady state as
\begin{equation}
 \frac 1{\vmoy(f)} = 
 \frac {\gamma}T \int_0^{+\infty} dy_a
 \Big\langle \exp\big\{{-\frac 1 T \big[\Vf(y_b)-\Vf(y_a+y_b)\big]}\big\}\Big\rangle_{\overleftrightarrow {y_b}}
 \label{eq:v_LD-V_S}
\end{equation}
where $\langle \ldots \rangle_{\overleftrightarrow y} $
denotes the translational average 
$\langle \mathcal O(y) \rangle_{\overleftrightarrow y} = \lim_{Y\to\infty}  \frac 1Y \int_0^Ydy\,\mathcal O(y)$,
and $\Vf(y)=V(y)-fy$ is the tilted potential. This expression is valid for an arbitrary potential~$V$, and the translational average in~\eqref{eq:v_LD-V_S} is expected to play the role of an averaging over $V$, for a random potential~$V$ of distribution invariant by translation along direction $y$.

Moreover, Gorokhov and Blatter~\cite{gorokhov_diffusion_1998} have given an elucidation
of the relation~\eqref{eq:v_LD-V_S} in terms of a mean first-passage time (MFPT) problem that we detail here, as it lies at the basis of the determination of the creep law for the interface that we propose in Sec.~\ref{sec:creep}.
Consider a particle starting from $y=0$ and denote by $\tau_1(Y)$ the MFPT
of the particle at a point $Y>0$ (we assume $f>0$ so that
the particle drifts towards positive $y$).
In a given potential $V$, the MFPT $\tau_1(Y;V)$
can again be determined exactly~\cite{gorokhov_diffusion_1998,hanggi_reaction-rate_1990} (see Fig.~\ref{fig:landscape_0D}) and reads
\begin{align}
 \tau_1(Y;V) &= \frac \gamma T
 \int_0^Y d y'
 \int_{-\infty}^{y'} d y''\, \ee^{-\frac{1}{T}  \left[\Vf(y'')-\Vf(y')\right]} 
 \label{eq:FPT_Boltzmann0}
\\
&= \frac \gamma T
 \int_0^Y d y' \int_{-\infty}^{y'} d y''\, 
 \frac{Z_V^f(y'')}{Z_V^f(y')}
 \label{eq:FPT_Boltzmann}
\end{align}
where $Z_V^f(y)=\ee^{-\frac{1}{T}  \Vf(y)}$ is the tilted Boltzmann-Gibbs
weight of the configuration~$y$.  Note that interestingly, $\tau_1(Y;V)$
is expressed in~\eqref{eq:FPT_Boltzmann} by means of the tilted weight $Z_V^f(y)$, although the non-equilibrium steady state of the particle is \emph{not}
proportional to this weight.  
The expression~\eqref{eq:FPT_Boltzmann} is exact, as the solution of the Pontryagin equation~\cite{pontryagin_disordered_2012} verified by the MFPT $\tau_1(Y;V)$ (see~\cite{gorokhov_diffusion_1998} for a pedagogical exposition).
We observe that in the low-temperature limit ${T} \to 0$, the form~\eqref{eq:FPT_Boltzmann0} yields by saddle-point asymptotics the expected Arrhenius (\emph{i.e.}~Kramers) behaviour of the MFPT by selecting the highest barrier of the tilted potential $\Vf$ situated before the arrival point $Y$:
\begin{align}
 \tau_1(Y;V) &
\ \underset{T\to 0}{\sim}\ 
  \ee^{\frac{1}{T}  \Delta \Vf (Y)} 
\qquad
 \Delta \Vf (Y) = 
\!\!\!\!\!\!
 \max_{\substack
  {
  \phantom{-.} 0<y'<Y
  \phantom{-.}\\
  -\infty < y'' <y'
  \phantom{-.}
  }
}
\!\!\!\!\!\!
\left[\Vf(y')-\Vf(y'')\right]
\label{eq:restau1_0d}
\end{align}
We emphasise however that the expression of the precise pre-exponential factor is non-trivially depending on the non-equilibrium nature of the steady-state~\cite{hanggi_reaction-rate_1990,bouchet_generalisation_2016}.
Nevertheless, an advantage of the MFPT approach is that the expression~\eqref{eq:restau1_0d}, in the low temperature limit and at exponential order, \emph{would be the same as in equilibrium settings}.
This observation proves useful below when extending the study to systems with a larger number of degrees of freedom.

The relations~(\ref{eq:FPT_Boltzmann0}-\ref{eq:restau1_0d}) were obtained for an arbitrary potential~$V$.
We now assume that the potential $V$ is a disorder which verifies a translational invariance in distribution.
Averaging the expression~\eqref{eq:FPT_Boltzmann} over disorder and separating the contribution of the drive~$f$, one gets:
\begin{align}
 \overline{\tau_1(Y;V)} &= \frac{\gamma}{T}
 \int_0^Y \dd y'\, \int_{-\infty}^{y'} \dd y''\;
 \ee^{\frac{1}{T}  f (y''-y')} \, 
 \overbrace{\,\overline{\ee^{-\frac{1}{T}  [V(y'')-V(y')]} }\,}^{\makebox[0pt]{\text{\scriptsize{depends only on $y\equiv y''-y'$}}}}
\\
 &= \frac{\gamma}{T}
 \int_0^Y \dd y'\, \int_{-\infty}^{0} \dd y\;
 \ee^{\frac{1}{T}  f y} \: 
 \overline{
 \ee^{-\frac{1}{T}  [V({y'+y})-V(y')]} 
 }
\\
 &= \frac{\gamma}{T}
   \int_0^Y \dd y'\, 
 \int_{-\infty}^{0} \dd y\;
 \ee^{ \frac{1}{T}  f y} \:
 \overline{
 \ee^{-\frac{1}{T}  [V({y})-V(0)]} 
 }
\\
 &= \frac{\gamma}{T}
 Y
 \int_{-\infty}^{0} \dd y\;
 \overline{
 \ee^{-\frac{1}{T}  [\Vf({y})-\Vf(0)]} 
 }
\label{eq:vmoypropY0d}
\end{align}
One thus obtains that, as expected, the average MFPT
$\overline{\tau_1(Y;V)}$ is proportional to the length $Y$ of the
interval to travel. This allows to define
consistently the mean velocity from
$\vmoy(f)=\frac{Y}{\:\overline{\tau_1(y;V)}\:}$ as follows
\begin{equation}
 \frac {1}{\vmoy(f)} =  \frac{\gamma}{T}
\int_{-\infty}^{0} \dd y\;
 \overline{
  \ee^{-\frac{1}{T}  [\Vf({y})-\Vf(0)]} 
 }
\end{equation}
One recovers the expression~\eqref{eq:v_LD-V_S} when the distribution of the disorder is invariant by translation along direction~$y$.

The approach using the exact solutions~\eqref{eq:v_LD-V_S} or~\eqref{eq:FPT_Boltzmann0} has not been extended to systems with more than one degree of freedom; however, the reasoning leading to the low-temperature limit~\eqref{eq:restau1_0d} can be adapted to systems with more degrees of freedom, as we detail in the next sections. Especially useful is the fact that such low-temperature approaches can be handled in or out of equilibrium by the use of a saddle-point analysis.

\section{The creep law from an effective description of the driven interface}
\label{sec:creep}

We design and study in this section an effective model aimed at capturing the behaviour at small force of the driven interface, reducing for a fixed length $\tf$ its infinite number of degrees of freedom to \emph{only two} degrees of freedom. %
The physical idea behind the effective model is that it allows to take into account \emph{(i)} the effects of elasticity and \emph{(ii)} the quasi one-dimensional motion of the interface centre of mass, along two \emph{orthogonal reduced coordinates}.
By comparing its behaviour for all available lengths~$\tf$, and optimising over $\tf$, we obtain by scaling its velocity-force dependence in a creep law form.
We define the model in Sec.~\ref{ssec:effectivemodel}, study in Sec.~\ref{ssec:MFPTeff} how the MFPT procedure developed in the previous section for one degree of freedom generalises to two degrees of freedom
. We study its scaling properties in Sec.~\ref{ssec:scalingandthecreeplaw} and derive the creep law for the effective model in Sec.~\ref{ssec:MFPT-to-creep} .

\subsection{Effective model}
\label{ssec:effectivemodel}

We focus on the problem of the non-equilibrium motion of the interface in a quasi\-static approximation: \emph{at fixed length $\tf$}, we assume that the extremities $\yi(\tau)$ and $\yf(\tau)$ of the interface follow a Langevin dynamics where the force derives from a potential given by the $f\neq 0$ \emph{equilibrium} point-to-point free energy
$
F^f_V(\tf,\yf|0,\yi)
$
defined in Sec.~\ref{ssec:tiltedDPf} in Eq.~\eqref{eq:defFfvinit}.
In this approach, the dynamics of the interface is thus reduced to the dynamics of a ``particle'' of coordinates given by the extremities $(\yi,\yf)$ of the interface
\begin{align}
\tilde\gamma \partial_\tau \yi(\tau)
& \ = \
-\partial_{\yi}F^f_V(\tf,\yf|0,\yi) + \sqrt{2\tilde\gamma\,\tilde T}\,\tilde\eta_\ii(\tau)
\label{eq:effdyni}
\\
\tilde\gamma \partial_\tau \yf(\tau)
& \ = \
-\partial_{\yf}F^f_V(\tf,\yf|0,\yi) + \sqrt{2\tilde\gamma\,\tilde T}\,\tilde\eta_\ff(\tau)
\label{eq:effdynf}
\end{align}
with effective friction~$\tilde \gamma$ and temperature $\tilde T$. Here the noises $\tilde\eta_\ii(\tau)$ and $\tilde\eta_\ff(\tau)$ are assumed to be independent Gaussian white noises of unit variance. They contribute to the equations of motion as thermal noises of effective temperature $\tilde T$.

\begin{figure}[t]
\centering  \includegraphics[width=0.95\textwidth]{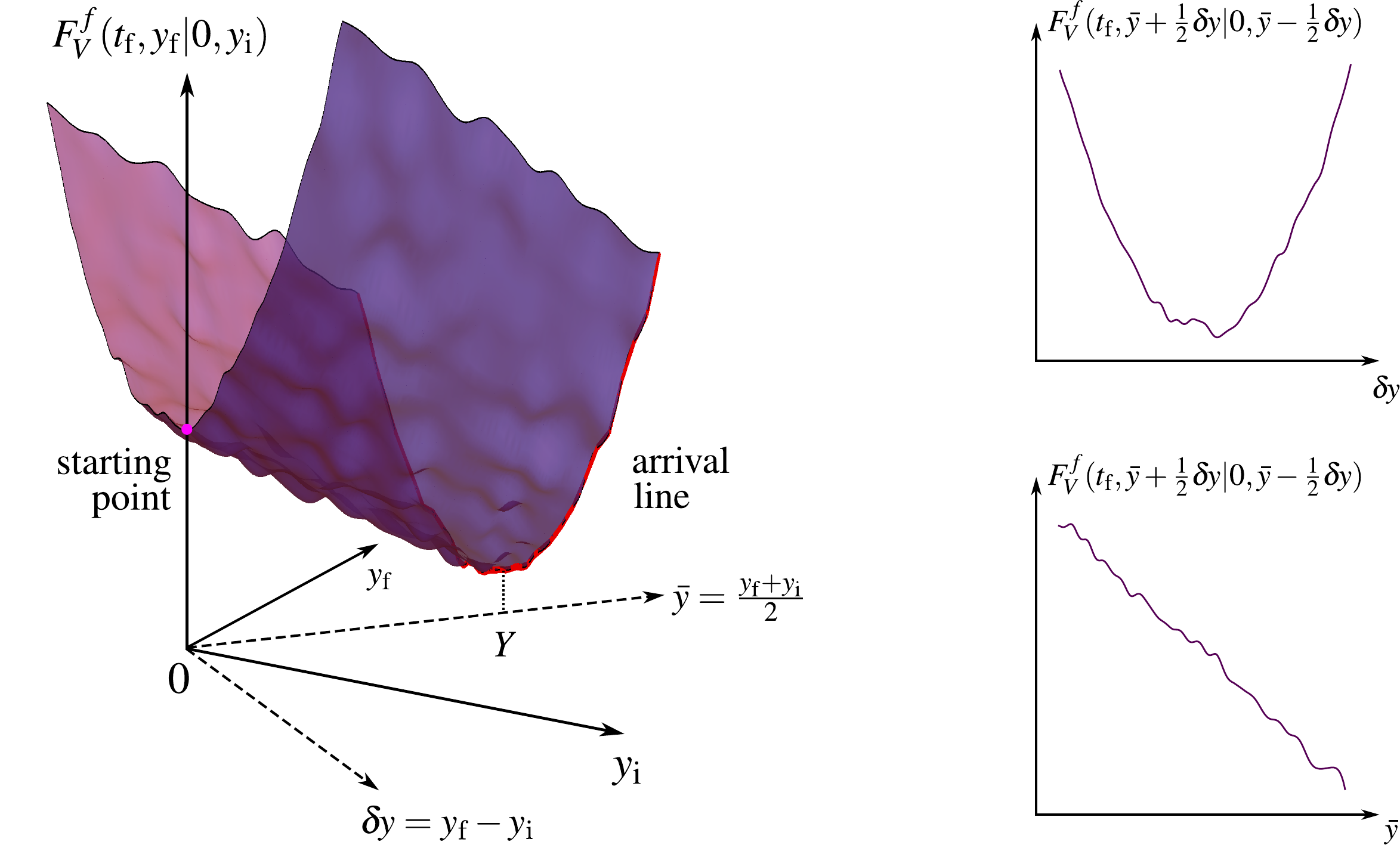}
\caption{
(\textbf{Left})
Schematic representation (at large~$\xi$)  of the effective potential $F_V^f(\tf,\yf|0,\yi)$ seen by the two extremities $\yi(\tau)$ and $\yf(\tau)$ in their quasi-static effective Langevin dynamics~(\ref{eq:effdyni}-\ref{eq:effdynf}), at \emph{fixed lengthscale} $\tf=\yf(\tau)-\yi(\tau)$.
As seen from~\eqref{eq:STSwithf_forMFPT}, the effective potential is a disordered tilted parabola, which drives the extremities along the direction of growing centre of mass $\bar y = \frac{\yf+\yi}{2}$ while the distance $ \deltay = {\yf-\yi}$ remains confined.
The effective motion of the two degrees of freedom $\yi(\tau)$ and $\yf(\tau)$ is thus well described by a quasi one-dimensional motion along direction $\bar y$.
As explained in Sec.~\ref{ssec:MFPTeff}, the evaluation of the mean velocity~$\vmoy(f)$ of the interface is translated into finding the MFPT between the starting point $(0,0)$ (purple dot) and the arrival line defined by $\bar y=Y$ (red line). In the low temperature regime, it is governed by the `instanton' trajectory (or `reaction path') joining this starting point and the arrival line, and which runs at the bottom of the tilted parabola.
As described in Sec.~\ref{ssec:scalingandthecreeplaw}, the evaluation of~$\vmoy(f)$ is performed by optimising over $\tf$: this implies that the effective potential is meaningful w.r.t.~the creep problem only for a specific \emph{force-dependent value} $\tf^\star(f)$ of the interface length, which induces an $f$-dependent effective potential.
(\textbf{Right})
Cuts of the  effective potential at fixed $\bar y$ as a function of $\deltay$ (top) and at fixed $\deltay$ as a function of $\bar y$ (bottom).
}
\label{fig:schemepoteff}
\end{figure}

We expect this approximation, where the interface extremities follow an overdamped gradient dynamics with thermal noise, to be valid in the limit of small mean velocity~$\vmoy(f)$, hence of small force~$f$. The underlying quasistatic hypothesis is that the global motion of the original interface is slow enough for the distribution of its extremities to remain well approximated by the $f\neq 0$ equilibrium one. Such model reduction is expected to be valid when there is a large time-scale separation between slow degrees of freedom (governing the average motion) and fast degrees of freedom (describing short-living fluctuations), a separation which one expects to be present for the driven 1D interface in the ${f\to 0}$ asymptotics {(equivalent to the ${\vmoy(f) \to 0}$ asymptotics)}.
If this holds, then the effective dynamics~(\ref{eq:effdyni}-\ref{eq:effdynf}) is a good candidate to determine $\vmoy(f)$ since it indeed possesses as a steady state the $f\neq 0$ equilibrium one of free energy given by $F_V^f(\tf,\yf|0,\yi)$.
Furthermore, even if the steady-state of the effective dynamics has zero mean velocity, the velocity of the interface can be estimated from the determination of a MFPT, irrespective of whether the steady state is in or out of equilibrium, as we argued in Sec.~\ref{sec:particle} for the particle and as we generalise in Sec.~\ref{ssec:MFPTeff}.
The MFPT will prove much easier to analyse for our effective model with two degrees of freedom than for the full 1D interface.

The effective temperature $\tilde T$ and friction $\tilde\gamma$, which fix the time scale, can be determined by comparing the full dynamics~\eqref{eq:Langevin_yttau_forcef} to the effective one~(\ref{eq:effdyni}-\ref{eq:effdynf}), in the absence of disorder (see Appendix~\ref{app:effectiveTgamma}).
One finds
\begin{equation}
  \tilde T \ =\ T \quad \qquad \tilde\gamma \ = \ \frac 12 \tf\,\gamma
  \label{eq:Ttildegammatilde}
\end{equation}
The second relation yields the correct dimension for $\tilde\gamma$ and expresses that the damping of the polymer endpoints dynamics grows with their separation $\tf$,  as physically expected for a segment of the interface of length~$\tf$ driven in its disordered environment. Note also that in this situation without disorder and for the choice~\eqref{eq:Ttildegammatilde} for the effective parameters, the effective model is exact in the sense that it yields the same steady-state distribution for the extremities as the complete model, as we detail in Appendix~\ref{app:effectiveTgamma}.

\subsection{Mean First Passage Time (MFPT)}
\label{ssec:MFPTeff}

In Sec.~\ref{sec:particle} for the single particle in a disordered one-dimensional potential, we could deduce the expression of its mean velocity from the exact expression of the MFPT for the Langevin equation~\eqref{eq:langevin0D}, which has one degree of freedom.
Here we are considering the effective coupled Langevin dynamics~(\ref{eq:effdyni}-\ref{eq:effdynf}) describing two degrees of freedom coupled by the effective potential $F^f_V(\tf,\yf|0,\yi)$. In that case, no exact expression is available, and we have to rely on a low-temperature asymptotics to estimate the MFPT.
To do so, one starts from a novel decomposition of the free energy, obtained in Appendix~\ref{sec:app_STS} by use of the Statistical Tilt Symmetry (STS) verified at non-zero force: we read from~\eqref{eq:STSwithf} that
\begin{equation}
  F_V^f(\tf,\yf|0,\yi) = \frac{c}{2\tf}(\yf-\yi)^2 -\frac {f\tf}2  (\yf+\yi)  + \bar F^f_V(\tf,\yf|0,\yi) +\text{const}(\tf)
  \label{eq:STSwithf_forMFPT}
\end{equation}
The first term acts as a confining potential which ensures that $\yi(\tau)$ and $\yf(\tau)$ remain close.
The second term drives the centre of mass $\frac12[\yf(\tau)+\yi(\tau)]$ at non-zero velocity, and establishes that the driving force $f$ affects the free energy by a linear contribution.
The third term, statistically invariant by a common translation of $\yi$ and $\yf$ (see Appendix~\ref{sec:app_STS}), plays the role of an \emph{effective disorder}~\cite{agoritsas_static_2013,agoritsas_temperature-dependence_2013}.
The last constant term, independent of $\yi$ and $\yf$, plays no role in the motion.
We thus argue that, in the long-time limit $\tau\to\infty$, the dynamics of this effective system is quasi one-dimensional along the direction of growing centre of mass, at the  bottom of a tilted parabolic potential (as illustrated on Fig.~\ref{fig:schemepoteff}).

From this insight, we can now extend to the interface the reasoning exposed in Sec.~\ref{sec:particle} for the particle. We first remark that different choices of boundary conditions for the interface will lead in the long-time limit either to an equilibrium steady state (\emph{e.g.}~in presence of confining walls) or to a non-equilibrium one (for periodic or free boundaries);
Nonetheless,  in those two settings, the MFPT problems have the same exponential behaviour in the low-temperature limit as long as they are dominated by trajectories which remain far from the possible boundaries.
This observation gives support to the use, for the study of the interface non-equilibrium velocity, of the $f\neq 0$ equilibrium free-energy~\eqref{eq:STSwithf_forMFPT} defined in Sec.~\ref{ssec:tiltedDPf}.

Let us now specify in detail the effective MFPT problem that we will analyse in order to evaluate $\vmoy(f)$.
The procedure consists in  evaluating the typical duration taken by the interface to travel a distance~$Y$ in a fixed disorder $V$.
The statistical invariance of $V$ along the longitudinal direction ($t$ axis) ensures that all points run at the same average velocity, so that the average velocity of an interface segment of length~$\tf$ is given by the velocity of its extremities $\yf$ and $\yi$.
In the effective description, one thus has to evaluate
the $\tf$-dependent MFPT $\tau_1(\tf,Y;V)$ to start from $(\yi,\yf)=(0,0)$ and to arrive on a line determined by $\frac 12 (\yi+\yf)=Y$, with $Y>0$ (see Fig.~\ref{fig:schemepoteff}), in the disorder-dependent effective potential~\eqref{eq:STSwithf_forMFPT}.
As in the case of the particle (Sec.~\ref{sec:particle}), the velocity at a scale $\tf$ will be evaluated as the average over the disorder of $Y/\tau_1(\tf,Y;V)$.

The MFPT $\tau_1(\tf,Y;V)$ is governed by the dominant trajectory (or `instanton', or `reaction path') between the starting point and the arrival line (Fig.~\ref{fig:schemepoteff}),
that arises, in the low-temperature regime, from the weak-noise Freidlin--Wentzell theory~\cite{freidlin_random_1998}, or from the equivalent WKB (Wentzel--Kramers--Brillouin) semi-classical theory~\cite{hanggi_reaction-rate_1990}, all yielding Kramer's escape rate~\cite{kampen_stochastic_2007}.
Since the effective potential confines the trajectories at the bottom of the parabola depicted in Fig.~\ref{fig:schemepoteff}, we argue that this instanton is quasi one dimensional.
It controls the MFPT $\tau_1(\tf,Y;V)$ by the Arrhenius time 
which
is obtained at dominant order in the low-temperature limit from the largest barrier to cross along the instanton. 
The natural coordinates are the centre of mass $\bar y$ and the endpoint difference~$\deltay$, defined as
\begin{equation}
  \bar y = \frac{\yf+\yi}{2}
\qquad\qquad
  \deltay = {\yf-\yi}
\end{equation}
The largest barrier along the {instanton} starts from a minimum $(\bar y'',\deltay'')$ and ends in a saddle point $(y',\deltay')$ of~\eqref{eq:STSwithf_forMFPT} at the top of the barrier (the top of the barrier is located at local maximum along the instanton, which lies itself at a saddle of the effective potential).
{Generalising the MFPT expression for the particle \eqref{eq:restau1_0d} by replacing the difference of 1D potential ${\left[\Vf(y')-\Vf(y'')\right]}$ by the difference of effective potential \eqref{eq:STSwithf_forMFPT} between the 2D coordinates $(\bar y'',\deltay'')$ and $(y',\deltay')$,}
the passage time writes
\begin{align} 
 \tau_1(\tf,Y;V) \sim
\frac{\tilde\gamma}{T}
\,&\ee^
{
-\frac 1T
 \big\{
  \frac{c}{2\tf} \left[ (\deltay'')^2 -(\deltay')^2 \right]
-
  f \tf \left(\bar y''-\bar y'\right)
 \big\}
}
\nonumber
\\[-2mm]
\times
&\,\ee^
{
-\frac 1T
 \big\{
  \bar F_V^f\left(\tf,\bar y''+\deltay''/2|0,\bar y''-\deltay''/2\right)
-
  \bar F_V^f\left(\tf,\bar y'+\deltay'/2|0,\bar y'-\deltay'/2\right)
 \big\}
}
\label{eq:restau1_non-ave}
\end{align}
The first line contains the parabolic and tilted part of the potential~\eqref{eq:STSwithf_forMFPT} corresponding respectively to the elasticity and to the driving force. The second line is the effective disorder induced by the disordered free-energy $\bar F^f_V$, which, at fixed~$\tf$, is invariant by translation along direction $y$ (see Appendix~\ref{ssec:STS_nonzero-force}).
This expression is the starting point of our analysis of the creep law and will allow us to justify the scaling procedure used in the literature for its standard derivation.

The expression~\eqref{eq:restau1_non-ave} is valid only at the exponential order and cannot be used to identify a prefactor proportional to the distance $Y$, at least not in the same way as it was done for the particle in equation~\eqref{eq:vmoypropY0d} after averaging over disorder. However, since \emph{(i)} the effective MFPT problem is quasi one-dimensional and \emph{(ii)} the effective disorder $\bar F^f_V$ is invariant in distribution by translation along the transverse direction~$y$, one obtains immediately that the disorder average of $\tau_1(\tf,Y;V)$ is as expected proportional to $Y$ at large enough~$Y$. 
(For instance, for integer~$Y$, one can split the interval $[0,Y]$ into $Y$ segments of length~1 and remark that the dominant endpoint on the arrival line of every segment is close to $\deltay=0$. This transforms the global MFPT problem into $Y$ successive MFPT problems having the same disorder-averaged passage time thanks to \emph{(ii)}.)
We will thus denote simply by $\tau_1(\tf;V)$ the MFPT $\tau_1(\tf,1;V)$ for $Y=1$, to which one can thus restrict, and evaluate the velocity using~$1/\tau_1(f;V)$.

Hence, we consider the expression~\eqref{eq:restau1_non-ave} as an estimator of the inverse velocity and we proceed in the next Section to the study of its scaling properties and to the optimisation over the polymer length $\tf$.
Note that in terms of the coordinates $(\yi,\yf)$ describing the polymer extremities, one has
\begin{align} 
 \tau_1(\tf;V) \sim
\frac{\tilde \gamma}{T}
\,&\ee^
{
-\frac{1}{T} 
 \big\{
  \frac c{2\tf} \big[ (\yf''-\yi'')^2-(\yf'-\yi')^2\big]
-
  \frac 12 f \tf \big[ (\yf''+\yi'')-(\yf'+\yi')\big]
 \big\}
}
\nonumber
\\[-2mm]
\times
&
\,\ee^
{
-\frac{1}{T} 
 \big\{
  \bar F_V^f(\tf,\yf''|0,\yi'')
-
  \bar F_V^f(\tf,\yf'|0,\yi')
 \big\}
}
\label{eq:restau1_non-ave_originalval}
\end{align}
This expression can be interpreted as the MFPT for a passage problem for the original interface, assuming that the free-energy cost of the drive $f$ is proportional to $f$ (see Sec.~II.A.4 in Ref.~\cite{blatter_vortices_1994}), an hypothesis which is in fact justified in our approach by the $f$-STS derived in Appendix~\ref{sec:app_STS}.

\subsection{Scaling analysis of the tilted free-energy}
\label{ssec:scalingandthecreeplaw}

The procedure to analyse the velocity-force dependence is the following: at fixed drive $f$, one will identify the optimal length of interface $\tf$ that dominates the motion, starting from the MFPT expression~\eqref{eq:restau1_non-ave_originalval}. Since one performs an average over the disordered potential $V$, one has the freedom to rescale the directions $y$ and $t$ in order to facilitate the scaling analysis, for instance as
\begin{equation}
  y=a\,\hat y
\qquad
  t=b\,\hat t
\label{eq:ytab}
\end{equation}
As we now explain, there is a peculiar choice of the scaling parameters $a$ and $b$ that allows one to study the low force regime.
The elastic (parabolic) and the force contribution to~\eqref{eq:restau1_non-ave_originalval} rescale upon~\eqref{eq:ytab} in an obvious way. The rescaling of the disordered free-energy contribution $\bar F^f_V$ is more complex but it is precisely the key point to comprehend, since \emph{this contribution summarises the effect of fluctuations {due to disorder} at all scales smaller than}~$\tf$.

The large-$\tf$ limit controls the small-velocity regime as will later be checked self-consistently: one will  thus first analyse the scaling properties of the tilted free-energy~\eqref{eq:STSwithf_forMFPT} in this limit, {keeping track of the disorder correlation length $\xi$ introduced in~\eqref{eq:VVxi}.}
The most simple case is that of the uncorrelated disorder ($\xi=0$): in this case, the large-$\tf$ distribution of $\bar F^f_V$ is the same as a Brownian motion in direction~$y$,
{up to a cutoff ${y \sim \tf^{2/3}}$,}
and of amplitude $cD/T$ (see \cite{halpin-healy_kinetic_1995,huse_huse_1985} at $f=0$; see also Appendix.~\ref{ssec:STS_nonzero-force} where~\eqref{eq:FfVWVf0} gives the result at $f\neq 0$).
This means that upon the rescaling~\eqref{eq:ytab}, one has
\begin{equation}
  \bar F^f_V(\tf,\yf|0,\yi;\xi=0) \stackrel{\text{(d)}}= a^{\frac 12}\big(\tfrac{cD}{T}\big)^{\frac 12} \, \hat F^f_{\hat V}(\hat \tf,\hyf|0,\hyi ;\xi=0)
\qquad [\text{for } \tf\to\infty]
 \label{eq:rescaling_free-energy_FbarfV_xi0}
\end{equation}
where $\hat F^f_{\hat V}$ is the disorder free energy for a polymer with an elastic constant $c=1$, a temperature $T=1$ and a disorder~$\hat V$ of amplitude $D=1$.
However, and this is one of the main results of this article, the low-temperature asymptotics used in the creep analysis cannot merely emerge from the $\xi=0$ case.
Indeed, the amplitude $cD/T$ would diverge as $T\to 0$, rendering the MFPT analysis impossible.
Instead, one has to rely on a free-energy scaling analysis with a short-range correlated disorder (recalling the definition~\eqref{eq:VVxi}, this corresponds to $\xi>0$)~\cite{agoritsas_static_2013,nattermann_interface_1988,agoritsas_temperature-induced_2010,agoritsas_disordered_2012,agoritsas_temperature-dependence_2013}.
Then,  no exact result is known for the full distribution of $\bar F^f_V$ in the large-$\tf$ asymptotic, but it has been shown that upon the rescaling~\eqref{eq:ytab}, and for $|\yi-\yf|\gtrsim\xi$, one has, instead of~\eqref{eq:rescaling_free-energy_FbarfV_xi0}:
\begin{equation}
  \bar F^f_V(\tf,\yf|0,\yi;\xi) \stackrel{\text{(d)}}= a^{\frac 12}\tilde D^{\frac 12} \, \hat F^f_{\hat V}(\hat \tf,\hyf|0,\hyi ;\xi/a)
\qquad [\tf\to\infty]
 \label{eq:rescaling_free-energy_FbarfV}
\end{equation}
where $\tilde D$ is the ``amplitude'' of the disorder free energy {two-point correlator}~\cite{agoritsas_static_2013,agoritsas_disordered_2012,agoritsas_finite-temperature_2012} at large $\tf$.
There is no {exact} expression for $\tilde D$ as a function of the parameters $\{c,D,T,\xi\}$, but the high- and low-temperature asymptotics are known:
\begin{equation}
  \tilde D \ \stackrel{T\gg T_\cc}{\sim}\  \frac{cD}{T}
\qquad\quad
  \tilde D \ \stackrel{T\ll T_\cc}{\sim}\ \frac{cD}{T_\cc}
\qquad\quad
\text{with }
T_\cc=(\xi c D)^{\frac 13}
\label{eq:lowhighTregimes}
\end{equation}
Here, $T_\cc$ is a characteristic temperature that separates these two asymptotic regimes, and $\tilde D$ presents a smooth crossover between them, {predicted analytically using a variational scheme \cite{agoritsas_temperature-induced_2010,agoritsas_temperature-dependence_2013} and observed numerically~\cite{agoritsas_disordered_2012,agoritsas_staticnum_2013}.}
The meaning of the characteristic temperature $T_\cc$ is that it separates a regime $T\gg T_c$ where the role of $\xi$ can be mostly ignored from a regime $T\ll T_c$ where on the contrary $\xi$ plays a major role while the temperature $T$ disappears from the free energy distribution. 
The quantitative analysis of the crossover between the high- and low-temperature regimes~\eqref{eq:lowhighTregimes} is discussed in Sec.~\ref{ssec:tempescaling}.
For the moment, we will simply use the fact that the amplitude $\tilde D$ retains the dependence in $\xi$ of the disorder free-energy, which manifests itself at scales much larger than $\xi$ itself~\cite{agoritsas_static_2013,agoritsas_disordered_2012,agoritsas_finite-temperature_2012}.
This fact has been used in those references to analyse the static ($f=0$) fluctuations of the interface, and we analyse in this article its consequences for the driven interface ($f>0$).

The first step of the rescaling procedure is to fix $a$ in~\eqref{eq:ytab} so as to ensure that the elastic and the disorder free-energy contributions in~\eqref{eq:STSwithf_forMFPT} scale {with the same prefactor:} 
one finds
\begin{equation}
  a = \Big(\frac{\tilde D}{c^2}\Big)^{\frac 13} b^{\frac 23}
 \label{eq:rescaclassical}
\end{equation}
{implying the following scaling in distribution} 
\begin{equation}
  F^f_V(\tf,\yf|0,\yi;\xi) 
\stackrel{\text{(d)}}=
  \Big(\frac{\tilde D^2 b}{c}\Big)^{\frac 13}
\Big[
  \frac{(\hyf-\hyi)^2}{2\hat\tf}  + \hat F^f_{\hat V}(\hat\tf,\hyf|0,\hyi;\tfrac \xi a) 
\Big]
-  \Big(\frac{\tilde D}{c^2}\Big)^{\frac 13} b^{\frac 53}f\; \frac {\hat  \tf}2  (\hyf+\hyi) 
\label{eq:rescaledFfV_freeb}
\end{equation}
We observe that the KPZ roughness exponent $\zeta=\frac 23$ appears naturally through the rescaling~\eqref{eq:rescaclassical}, which matches by power counting the elastic and disorder scaling exponents~\cite{huse_huse_1985}. In fact, in the absence of driving force~$f$, the asymptotic expression of the roughness function of the interface can be inferred from a \emph{saddle-point argument} at large~$\tf$~\cite{agoritsas_static_2013} using the rescaling~\eqref{eq:rescaclassical} at~$f=0$. 
One main result of the present article is that this argument can be adapted and generalised to determine at $f>0$ the mean velocity of the driven interface. 

To do so, in the presence of the driving force~$f$, the second step of the scaling analysis is to choose the rescaling factor $b$ in~\eqref{eq:ytab} so as to match the prefactors of the force-dependent and the force-independent contributions to~\eqref{eq:rescaledFfV_freeb}: one finds
\begin{equation}
  b = (c \tilde D )^{\frac 14} f^{-\frac 34}
 \label{eq:rescbf}
\end{equation}
and this yields (still in the large $\tf$ limit):
\begin{equation}
  F^f_V(\tf,\yf|0,\yi;\xi) 
\stackrel{\text{(d)}}=
  f^{-\frac 14}
  \Big(\frac{\tilde D^3 }{c}\Big)^{\frac 14}
\Big[
  \frac{(\hyf-\hyi)^2}{2\hat\tf}  + \hat F^f_{\hat V}(\hat\tf,\hyf|0,\hyi;\tfrac \xi a) 
-
   \frac {\hat  \tf}2  (\hyf+\hyi) 
\Big]
\label{eq:scalingFfall}
\end{equation}
Upon this choice, one now has 
\begin{equation}
a=\Big(\frac{\tilde D}{cf}\Big)^{\frac 12}
\label{eq:adepf}
\end{equation}
as obtained from~\eqref{eq:rescaclassical} and~\eqref{eq:rescbf}.
The force dependence is in fact fully contained in the common prefactor~$\propto f^{-1/4}$ in the large~$\tf$ limit, as we now justify.
The explicit dependence in~$f$ of 
$\hat F^f_{\hat V}$ is removed upon disorder averaging, 
since~$f$ appears only through the transformed disorder~\eqref{eq:transfoVdepf}, which one can translate along direction~$y$ at any fixed $\tf$ in order to absorb the  dependence in~$f$ without changing its distribution {(see Appendix~\ref{ssec:STS_nonzero-force})}.
The last dependence in~$f$ is through the rescaled disordered correlation length $ \xi / a$ where $a$ is given by~\eqref{eq:adepf}; in the small $f$ regime, the rescaled correlation length $ \xi / a$ goes to zero since ${a\sim f^{-1/2}}$ and this dependence in $f$ can be dismissed as one thus recovers for $\hat F^f_{\hat V}$ an uncorrelated disorder. We emphasise that, however, the original correlation length~$\xi$ still remains present in the problem through the amplitude~$\tilde D$~\cite{agoritsas_static_2013,agoritsas_temperature-induced_2010}.

At low enough~$f$, we have thus obtained that the landscape of potential~\eqref{eq:scalingFfall} seen by the effective degrees of freedom $\yi$ and $\yf$, once rescaled by~\eqref{eq:rescaclassical} and~\eqref{eq:rescbf}, depends in the driving force only through its overall prefactor $f^{- 1/4}$ .
\emph{This means that the rescaled locations} $\hyfi'$ \emph{and~}$\hyfi''$ \emph{of the minimum and the saddle of the Arrhenius barrier of the MFPT problem exposed in Sec.~\ref{ssec:MFPTeff} do not depend on the force $f$}.
This property, explicitly constructed in our approach, justifies 
the often stated argument that all barriers of the creep problem scale in the same way in $\propto f^{-1/4}$.
In particular, this implies that if other barriers of height comparable to the dominant one contribute to the MFPT, they do not affect its scaling form.

We furthermore observe that, crucially, the rescaled effective potential~\eqref{eq:scalingFfall} is at a rescaled force equal to $1$, implying also that the backward barrier (between the saddle and the minimum reached after crossing the saddle) is much higher than the forward barrier, the non-rescaled difference being of order $\propto f^{-1/4}$. This justifies consistently that we neglect the backward motion in comparison to the forward motion in the evaluation of the velocity (see Fig.~\ref{fig:barriers-modif-creep}, left and Sec.~\ref{sec:finite-size} for a situation where the backward motion plays a role).

\subsection{Scaling of the mean first passage time and the creep law}
\label{ssec:MFPT-to-creep}

One can now evaluate the disorder-averaged MFPT $\bar\tau_1(f)$ of the interface by integrating~\eqref{eq:restau1_non-ave_originalval} over every possible {length~${\tf \in \left[0,L \right]}$ of an interface segment, assuming that ${L \gg \mathcal{L}_\cc}$ with $\mathcal{L}_\cc$ the lengthscale above which the Brownian rescaling~\eqref{eq:rescaling_free-energy_FbarfV} is valid ($\mathcal{L}_\cc$ will be identified later on as the `Larkin length', see Sec.~\ref{ssec:tempescaling})}. Using the rescaling~\eqref{eq:scalingFfall}, one has
\begin{align}
\!
 \bar\tau_1(f)
 \stackrel{(L > \mathcal{L}_\cc)}{=}
&
 \frac{1}{L}\int_0^{L} \!\!\!d\tf \;
 \overline{\tau_1(\tf;V)}
 = \frac{1}{L}\int_0^{\mathcal{L}_\cc} \!\!\!d\tf \;
 \overline{\tau_1(\tf;V)} + \frac{1}{L}\int_{\mathcal{L}_\cc}^L \!\!\!d\tf \;
 \overline{\tau_1(\tf;V)}
\\
 \stackrel{(L \gg \mathcal{L}_\cc)}{\sim}
&
\frac{1}{\hat{L}}\int_{\hat{\mathcal{L}}_\cc}^{\hat{L}} \!\!\!d\htf \;
\mathbb E_{\hat V}
\ee^
{
-\frac{1}{T} f^{-\frac 14}
 \big[\frac{\tilde D^3 }{c}\big]^{\frac 14}
 \Big\{
  \frac{(\hyf''-\hyi'')^2-(\hyf'-\hyi')^2}{2\htf}
-
  \htf \frac{(\hyf''+\hyi'')-(\hyf'+\hyi')}{2}
}
\nonumber
\\[-5mm]
&
\hspace*{.4\columnwidth}
\hfill
\phantom{e}^{
+
\
  \hat F_{\hat V}^f(\htf,\hyf''|0,\hyi'';\xi/a)
  -
  \hat F_{\hat V}^f(\htf,\hyf'|0,\hyi';\xi/a)
 \Big\}
}
\label{eq:vftimeintegral}
\end{align}
where $\mathbb E_{ \hat V}$ is another notation for the disorder average.
In this expression $\hyfi'$ and $\hyfi''$ are, as we discussed above, the $f$-independent but $\htf$-dependent locations of the extremities of the dominant barrier of the rescaled MFPT problem.
\emph{The form~\eqref{eq:vftimeintegral} of the velocity is thus amenable to a saddle-point estimation in the low-force limit} $f\to 0$:
in that limit, the integral is dominated by the maximum of the exponent.
Since all the dimensioned parameters have been factored out in this exponent, the saddle-point is reached at a value $\htf^\star$ of $\htf$ which is independent of the dimensioned parameters of the problem, and in particular independent of $f$.
We emphasise that this construction requires the system to be large enough (${L \gg \mathcal{L}_\cc}$) in order that
\textit{(i)}~the KPZ scaling \eqref{eq:rescaclassical} is meaningful for a large range of segment length $\tf$,
\textit{(ii)}~we can neglect the contribution of ${\tf \in [ 0,\mathcal{L}_\cc ]}$ in the MFPT,
so we can self-consistently assume that the saddle point is reached at  ${\htf^\star \in [\hat{\mathcal{L}}_\cc,\hat{L} ]}$.
We refer to Sec.~\ref{ssec:tempescaling} for the corresponding regime of validity.
%
We finally obtain
\begin{align}
 \bar\tau_1(f)
 \ \sim \
&
\mathbb E_{\hat V}
\ee^
{
-\frac{1}{T} f^{-\frac 14}
 \big[\frac{\tilde D^3 }{c}\big]^{\frac 14}
 \Big\{
  \frac{(\hyf''-\hyi'')^2-(\hyf'-\hyi')^2}{2\htf^\star}
-
  \htf^\star \frac{(\hyf''+\hyi'')-(\hyf'+\hyi')}{2}
}
\nonumber
\\[-5mm]
&
\hspace*{.4\columnwidth}
\hfill
\phantom{e}^{
+
\
  \hat F_{\hat V}^f(\htf^\star,\hyf''|0,\hyi'';\xi/a)
  -
  \hat F_{\hat V}^f(\htf^\star,\hyf'|0,\hyi';\xi/a)
 \Big\}
}
\label{eq:finaltau1}
\end{align}
where the $\hyfi'$ and $\hyfi''$ are evaluated in $\htf=\htf^\star$.
This last step of the scaling analysis justifies why a naive power-counting argument on the free energy yields the correct creep exponent $\mu=1/4$: it is because setting the three contributions of the free-energy to the same scale allows to perform a saddle-point analysis at $f\to 0$, as we have presented. 

The stretched exponential scaling in $f$ holds for all $V$. 
one, then the disorder average~\eqref{eq:finaltau1} gives the correct exponential behaviour for $\overline{1/\tau_1(f)}\sim 1/\overline{\tau_1(f)}$, hence yielding to the average velocity 
We assume that the distribution of MFPT is peaked enough in the effective model 
so that $\overline{1/\tau_1(f;V)}\sim 1/\bar\tau_1(f)$
(see Ref.~\cite{kolton_uniqueness_2013} for a study of self-averaging in the large size-limit
and 
see Ref.~\cite{malinin_transition_2010} for a generic study of the distribution of activation times).
One finally obtains the creep form of the mean velocity
\begin{equation}
  \vmoy(f) \,\sim\, 
\,
\ee^{\text{
$
\displaystyle{-\frac {U_\cc}T \Big(\frac{f_\cc}f\Big)^\frac14}\Delta \hat F^\star
$
}}
  \label{eq:creep_1d_TM}
\end{equation}
where $\Delta \hat F^\star$ is a numerical factor given by the (positive) adimensioned  difference of free-energy between the minimum and the saddle, evaluated at $\htf=\htf^\star$ as appearing in~\eqref{eq:finaltau1}.
One identifies the creep exponent $\mu=\frac 14$.
We emphasise here one important aspect of the creep phenomenology: the relation~\eqref{eq:creep_1d_TM} can be read as an corresponding to an Arrhenius waiting time with an effective barrier ${U_\cc}\;({f_\cc}/f)^{1/4}$ which \emph{diverges} as $f\to 0$. In our description, this arises from a fine-tuned rescaling of the MFPT where we used the distributional properties of the free energy (arising from those of the disorder), \emph{i.e.}~not reasoning at a fixed realisation of the disorder~$V$. 

By direct identification between~\eqref{eq:finaltau1} and~\eqref{eq:creep_1d_TM}, one reads that the characteristic energy $U_\cc$ and characteristic force $f_\cc$ are related by 
$ 
U_\cc^4 f_\cc = \tilde D^3/c 
$.
This relation does not fix their expressions. To do so, one can for instance impose that $f_\cc$ does not depend on temperature (which makes the analysis more simple, but the results do not depend on this arbitrary choice).
However, the temperature dependence of $\tilde D$ is not known exactly: it crosses over from a high-temperature to a low-temperature regime which are both well-understood (see Eq.~\eqref{eq:lowhighTregimes}), but the crossover itself has only be determined through variational and numerical approaches~\cite{agoritsas_static_2013,agoritsas_temperature-induced_2010}.
In the low-temperature regime $T\ll T_\cc$ we are interested in, one has $\tilde D\sim \frac{cD}{T_\cc}$
, thus implying
\begin{equation}
 U_\cc=T_\cc \stackrel{\eqref{eq:lowhighTregimes}}{=}(\xi c D)^{\frac 13}
 \:,\quad
 f_\cc
 =\frac{\tilde D^3}{c T_\cc^4}
 =\Big(\frac{D^2}{c \xi^7}\Big)^{\frac 13}
\qquad
[T\ll T_\cc]
\label{eq:UcfclowT}
\end{equation}

One also remarks that, coming back to the $f$-dependent variable $t$ instead of $f$-independent $\hat t^\star$ through the scalings~\eqref{eq:ytab} and~\eqref{eq:rescbf}
one obtains the optimal lengthscale at which the creep motion occurs:
\begin{equation}
  L_{\opt}(f) = (c \tilde D )^{\frac 14} f^{-\frac 34}
  \label{eq:Lopt_topt_f34}
\end{equation}
It diverges as $f\to 0$, justifying self-consistently the study of the large $\tf$ limit in order to understand the low-force regime.
{This scaling has been numerically verified in the recent study Ref.~\cite{ferrero_2016_ArXiv-1604.03726}, where the $L_\opt(f)$ is found to play the role of a cut-off length in the distribution of avalanche sizes in the motion of the driven interface.}

\section{Finite-size analysis and phase diagram of the flow and creep regimes}
\label{sec:finite-size}

We now detail how the effective-model approach can be used to extend the regime of forces that one can describe quantitatively from the standard creep regime to lower and larger forces.
{We first discuss the correction of the creep law when taking into account finite-size effects,} which are responsible of a crossover between the usual creep regime and a linear-response Ohmic regime at very small forces.
{Then we construct the phase diagram for the low temperature limit
and we discuss the additional temperature-dependent corrections.
Last, we discuss how the creep law can be modified at intermediate forces.}

\subsection{Derivation of the finite-size behaviour in the $f\to 0$ asymptotics}
\label{ssec:fsize}

Observing the form of the creep law~\eqref{eq:creep_1d_TM}, one remarks that the function $\vmoy(f)$ is non-analytic in zero: a Taylor expansion of $\vmoy(f)$ for $f$ around 0 yields 0 at any order. In particular, a linear response regime where $\vmoy(f)$ would be linear in~$f$ is absent.
As we now discuss, these features follow from the hypothesis made that the system is infinite. From the expression~\eqref{eq:Lopt_topt_f34} we indeed read that the optimal portion of interface $L_{\opt}(f)$ to move under the drive~$f$ can take any arbitrarily large value as the force goes to zero.

\begin{figure}
\centering 
 \includegraphics[width=0.42\columnwidth]{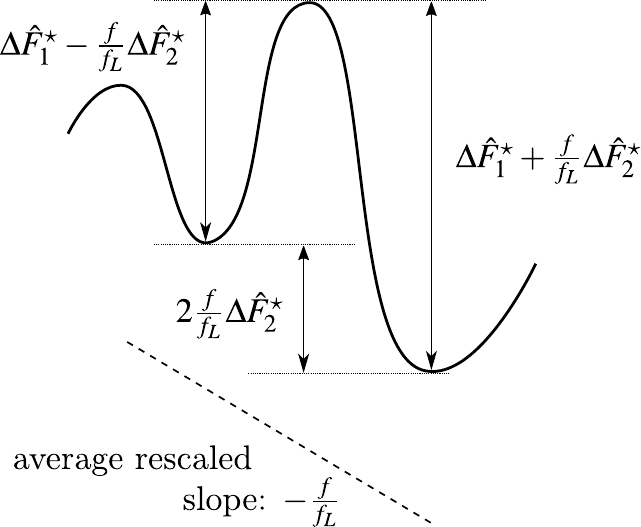}\hfill\includegraphics[width=0.55\columnwidth]{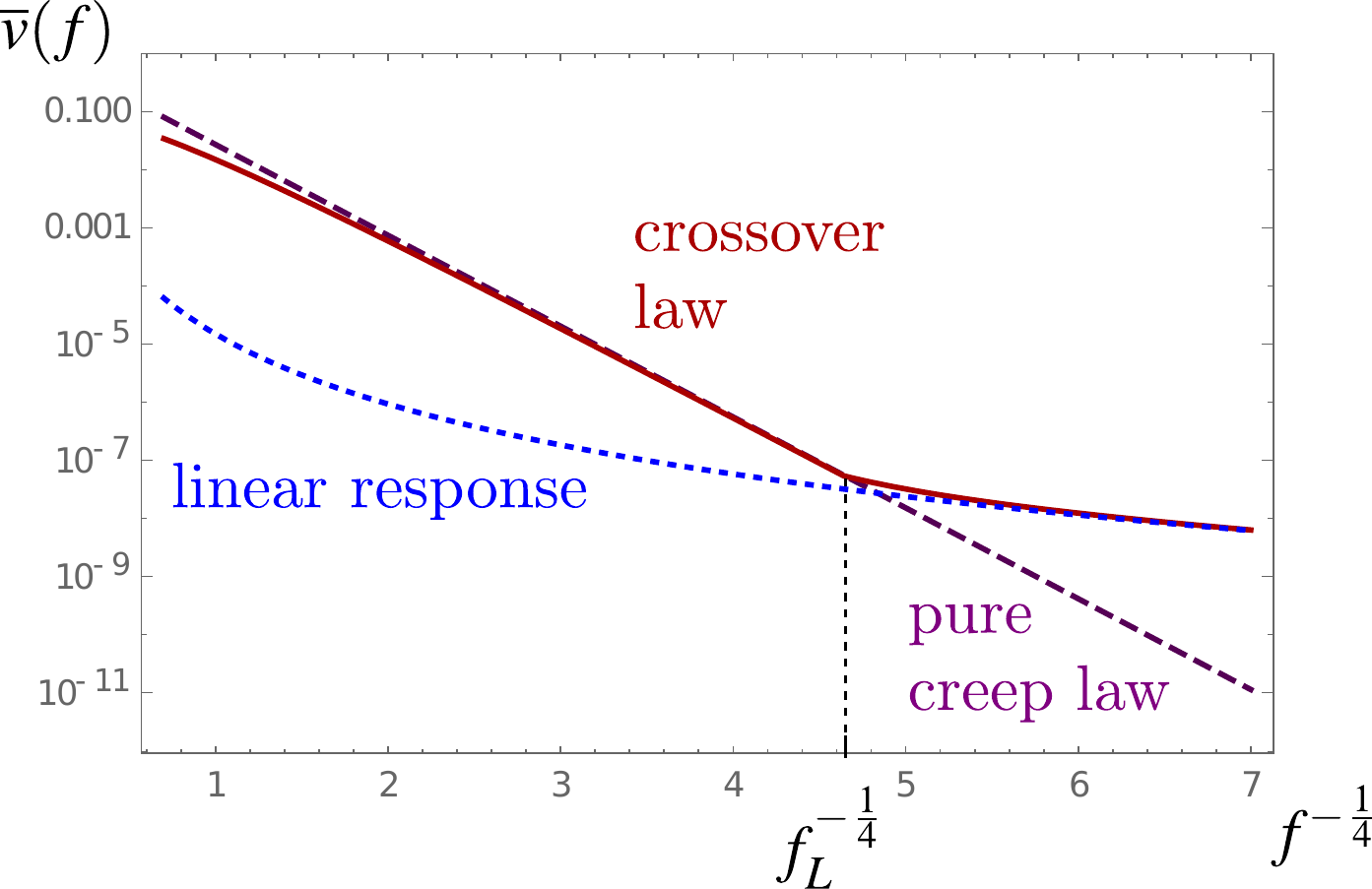} 
\caption{ (\textbf{Left})
Schematic representation of the rescaled free-energy barriers corresponding to the forward and backward motion in the effective model.
The contribution $\Delta \hat F^\star_1$ represents the elastic and disorder part of~\eqref{eq:rescaledFfV_bL}, while $\frac{f}{f_L}\Delta \hat F^\star_2$ represents the contribution arising from the driving force.
The scaling of those barriers is that of the low-force regime $0< f\lesssim f_L$, where the optimal length displaced by the driving force is of the order of the system size~$L$.
The difference between the forward and backward fluxes yields the expression~\eqref{eq:creep_1d_finiteL_sh} for the average velocity in this low-force regime.
(\textbf{Right})
Mean velocity $\vmoy(f)$ (in log scale) as a function of $f^{-1/4}$: in dashed purple, the infinite system-size creep law~\eqref{eq:creep_1d_TM}; in continuous red line, the extended creep law~\eqref{eq:creep_1d_finiteL_sh} crossing between the pure stretched exponential behaviour~\eqref{eq:creep_1d_TM} and the linear behaviour, expansion of~\eqref{eq:creep_1d_finiteL_sh} at small forces, represented in blue dotted line.
The bending of the extended creep law $\vmoy(f)$ (purple continuous line) at small values of $f^{-1/4}$ corresponds to having taken into account the backward flux (second line of~\eqref{eq:creep_1d_finiteL_sh}).
Parameters are $\tilde D=1$, $c=1$, $T=1$, $L=100$, $\Delta F_1^\star=4$, $\Delta F_2^\star=0.4$.
}
\label{fig:barriers-modif-creep}
\end{figure}

For a finite system of size~$L$, this can of course not hold.
In fact, for low enough forces $f\lesssim f_L$ with $f_L$ defined as
\begin{equation}
  L_\opt(f_L)\equiv L
  \qquad\text{\emph{i.e.}}\qquad
  f_L = (c \tilde D )^{\frac 13}\,L^{-\frac 43}
  \label{eq:eqfL}
\end{equation}
the saddle-point asymptotics that we have derived in the previous section is not valid anymore, because the solution $\htf^\star$ would correspond to an optimal portion of the interface $L_\opt(f)$ larger than the system size~$L$.
To evaluate correctly the mean velocity~$\vmoy(f)$, one first has to take into account that the rescaling parameter $b$ in~\eqref{eq:rescaledFfV_freeb} saturates to $b=L$ for $f<f_L$, instead of taking the value $b=L_\opt(f)$ as in~\eqref{eq:rescbf}.
A second change to take into account is that, as a consequence, the barriers of the problem do not all scale in the same way anymore, meaning that the mean slope of the rescaled landscape of potential is not equal to one as in~\eqref{eq:scalingFfall}: instead, one now has 
\begin{equation}
  F^f_V(\tf,\yf|0,\yi;\xi) 
\stackrel{\text{(d)}}=
  \Big(\frac{\tilde D^2 L}{c}\Big)^{\frac 13}
\bigg\{
\Big[
  \frac{(\hyf-\hyi)^2}{2\hat\tf}  + \hat F^f_{\hat V}(\hat\tf,\hyf|0,\hyi;\tfrac \xi a) 
\Big]
-  \frac{f}{f_L}
\; \frac {\hat  \tf}2  (\hyf+\hyi) 
\bigg\}
\label{eq:rescaledFfV_bL}
\end{equation}
with $a = (\frac{\tilde D}{c^2})^{\frac 13} L^{\frac 23}$ as read from~\eqref{eq:rescaclassical}.
The rescaled force thus reads~$\frac{f}{f_L}$ instead of~$1$ as in~\eqref{eq:scalingFfall}; hence, at fixed system size $L$, the rescaled force goes to zero as $f$ goes to zero, meaning that one cannot neglect the backward motion as we did in Sec.~\ref{ssec:scalingandthecreeplaw}.
The mean velocity can then be estimated by the flux difference between the forward and backward inverse average waiting times.
Denoting as on Fig.~\ref{fig:barriers-modif-creep} the rescaled free-energy barrier of the forward (resp. backward) motion by $\Delta\hat F^\star_1-\frac{f}{f_L}\Delta \hat F^\star_2$ (resp. $\Delta\hat F^\star_1+\frac{f}{f_L}\Delta \hat F^\star_2$), one gets
\begin{equation}
  \vmoy(f) \,\approx\, 
\ee^{\text{
$
\displaystyle{
-\frac 1T 
\Big(\frac{\tilde D^2 L}{c}\Big)^{\frac 13}
\Big[\Delta F^\star_1 - \tfrac{f}{f_L}  \Delta F^\star_2\Big]
}
$
}}
-
\,
\ee^{\text{
$
\displaystyle{
-\frac 1T 
\Big(\frac{\tilde D^2 L}{c}\Big)^{\frac 13}
\Big[\Delta F^\star_1 + \tfrac{f}{f_L}  \Delta F^\star_2\Big]
}
$
}}
  \label{eq:creep_1d_finiteL}
\end{equation}
{for forces ${f<f_L}$, whereas at ${f >f_L}$ taking into account the backward contributions modifies  \eqref{eq:creep_1d_TM} as follows (with ${\Delta \hat F^\star=\Delta F^\star_1 -  \Delta F^\star_2}$):
\begin{equation}
  \vmoy(f) \,\approx\, 
\ee^{\text{
$
\displaystyle{
-\frac 1T 
\Big(\frac{\tilde D^2 L_\opt(f)}{c}\Big)^{\frac 13}
\Big[\Delta F^\star_1 -  \Delta F^\star_2\Big]
}
$
}}
-
\,
\ee^{\text{
$
\displaystyle{
-\frac 1T 
\Big(\frac{\tilde D^2 L_\opt(f)}{c}\Big)^{\frac 13}
\Big[\Delta F^\star_1 + \Delta F^\star_2\Big]
}
$
}}
  \label{eq:creep_1d_finiteL-bis}
\end{equation}
This finally yields the predictions:
}
\begin{equation}
    \vmoy(f) \,\approx\, 
    \begin{cases}
      \ee^{\text{
          $ \displaystyle{ -\frac 1T \Big(\frac{\tilde D^2
              L}{c}\Big)^{\frac 13} \Delta F^\star_1 } $
        }} \sinh \Big[ \frac 1T \frac{f}{f_L} \Big(\frac{\tilde D^2
        L}{c}\Big)^{\frac 13} \Delta F^\star_2 \Big]
&\quad
(f\lesssim f_L)
\\
\ee^{\text{
    $ \displaystyle{ -\frac 1T f^{-\frac 14} \Big(\frac{\tilde D^3
        }{c}\Big)^{\frac 14} \Delta F^\star_1 } $
}}
 \sinh \Big[ \frac 1T f^{-\frac 14} \Big(\frac{\tilde D^3
        }{c}\Big)^{\frac 14} \Delta F^\star_2 \Big]
&\quad
(f\gtrsim f_L)
    \end{cases}
  \label{eq:creep_1d_finiteL_sh}
\end{equation}
The second line takes into account the backwards flow and yields at dominant order the creep law~\eqref{eq:creep_1d_TM} in the low-$f$ regime.
We illustrate on Fig.~\ref{fig:barriers-modif-creep}\,(right) how this extended creep law crosses over between the standard stretched exponential behaviour and the linear behaviour if $\vmoy(f)$ at small values of $f$ that arises from an expansion of~\eqref{eq:creep_1d_finiteL_sh} around $f=0$.
Note that in~\eqref{eq:creep_1d_finiteL} we have not made explicit the pre-exponential factors which are the same for the two terms of the difference, as they stem from the same barrier (see Fig.~\ref{fig:barriers-modif-creep}, left). This ensures in particular that the $f\to 0$ limit of~\eqref{eq:creep_1d_finiteL_sh} is zero. Besides, the apparent singular behaviour is due to the abrupt transition between the $L_\opt=L$ and the $L_\opt\sim f^{-3/4}$ regimes as $f$ decreases, but is of course in fact smooth.

The generic form with a hyperbolic sine in the velocity-force behaviour \eqref{eq:creep_1d_finiteL_sh} is not new, as it is obtained by taking into account the forward \emph{and} backward Arrhenius contributions over an energy barrier when quantifying such a thermally-assisted flux flow (TAFF) \cite{anderson_kim_1964_RevModPhys36_39}.
Such a crossover from creep to flow at very small driving forces has actually been observed experimentally in Ref.~\cite{kim_interdimensional_2009} by Kim and coauthors for ferromagnetic domain walls;
within the interpretation of a dimensional crossover, the domain wall was treated in the flow regime as a particle in 1D random potential
(see also Ref.~\cite{leliaert_creep_2016} for the influence of system size on conductivity).
%
The novelty that we bring is thus twofold:
on the one hand we derive the prediction \eqref{eq:creep_1d_finiteL_sh} from our effective model, rather than assuming an \textit{ad hoc} 1D potential; 
on the other hand we obtain  the explicit scaling of its parameters with the constants of interface model, in particular in the low-temperature asymptotics, together with a study of its regime of validity as exposed in the following Subsections.
We mention furthermore than such a generic crossover has also implications for the phase diagram of type-II superconductors, as pointed out for instance in Ref.~\cite{kes_1989_SupercondSciTechnol1_242}, since magnetic vortices can be described as elastic lines embedded in a disordered environment (but a 3D embedding space, instead of the 2D one we are considering).


\begin{figure}[th]
  \centering
  \includegraphics[width=.39\columnwidth]{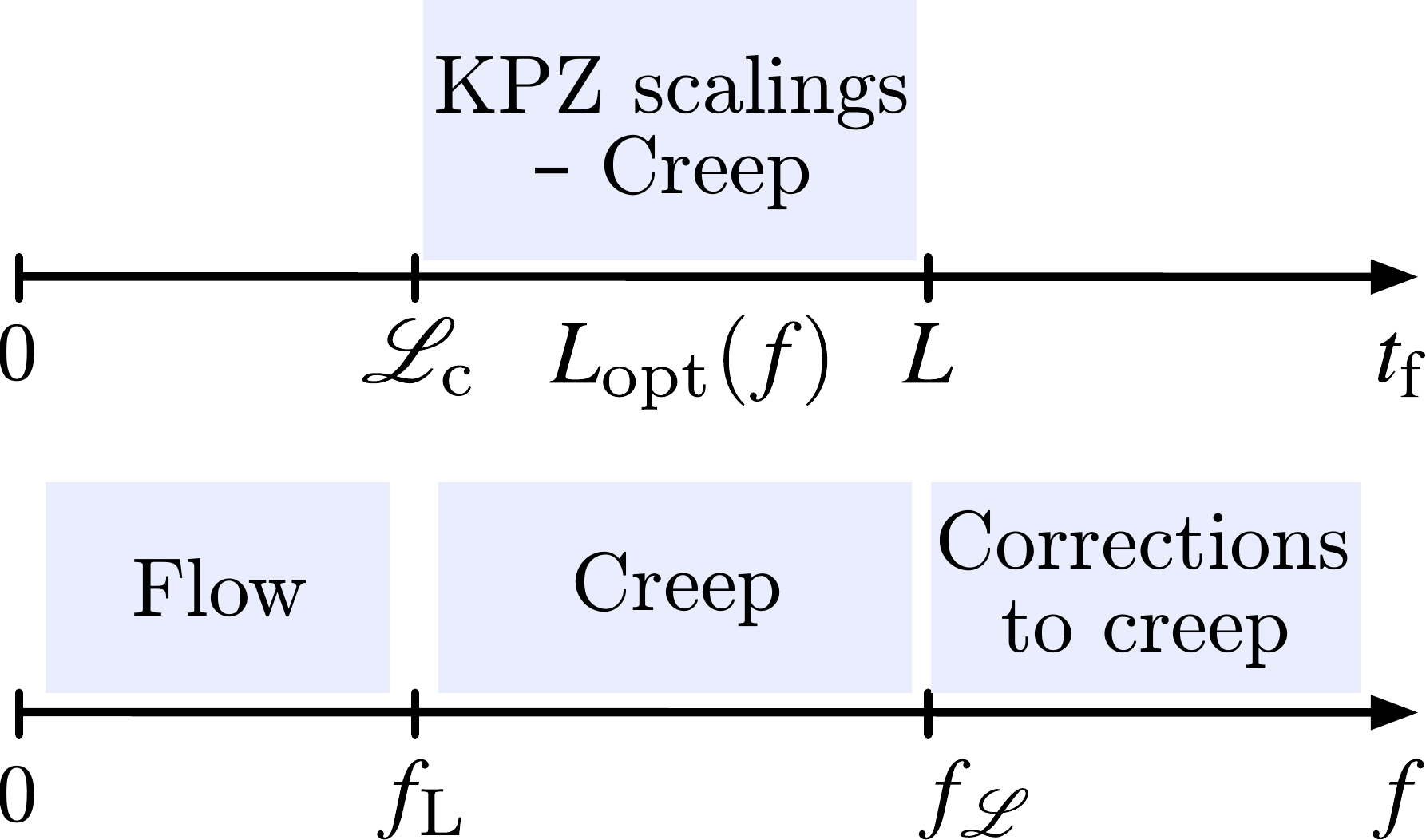}
\hfill
  \includegraphics[width=.55\columnwidth]{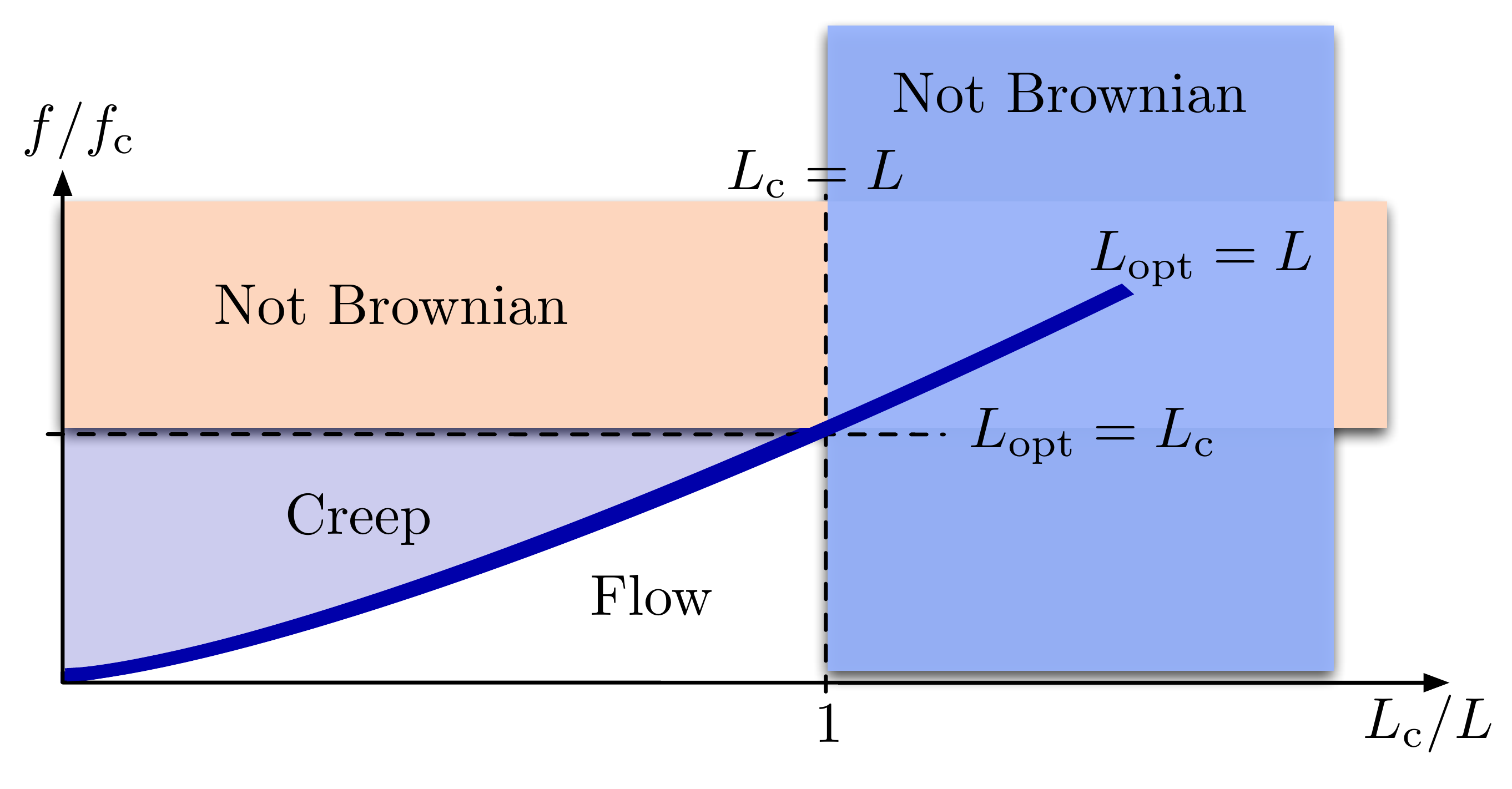}
\caption{
(\textbf{Left})
Illustration of the criteria determining the phase diagram: $L_\opt(f)$ has to be larger than the temperature-dependent Larkin length $\mathcal L_\cc$ and smaller than system size $L$ for the scaling {analysis} to yield the creep law. These two conditions are translated into a self-consistent condition $f_{\text{L}}< f < f_{\mathcal L}$ on the force for the creep regime to hold.
(\textbf{Right}) Corresponding phase diagram in the $(L_\cc/L,f/f_\cc)$ place in the $T\to 0$ asymptotics.
See Fig.~\ref{fig:phase-diagram} for the full diagram including the behaviour at $T>0$.
}
  \label{fig:2D_phase-diagram}
\end{figure}

\subsection{Creep and flow regimes in the zero-temperature asymptotics}
\label{ssec:phasediag}

Let us now establish the phase diagram represented on Fig.~\ref{fig:phase-diagram} and Fig.~\ref{fig:2D_phase-diagram}. We first focus in this subsection on the low-temperature asymptotics. The crossover line between the creep regime and the flow regime is given by the equation $L=L_\opt(f)$ (with $L_\opt(f)$ given by~\eqref{eq:Lopt_topt_f34}), that one can rewrite as
\begin{equation}
  L = (c\tilde D)^{\frac 14} (f_\cc)^{-\frac 34} \Big(\frac{f_\cc}{f}\Big)^{\frac 34} = \frac{T_\cc^5}{cD^2} \Big(\frac{f_\cc}{f}\Big)^{\frac 34} 
\end{equation}
where we used the low-temperature behaviour~\eqref{eq:UcfclowT} of $f_\cc=[D^2/(c \xi^7)]^{1/3}$ and of $\tilde D$.
This suggests to define a {typical lengthscale}:
\begin{equation}
L_\cc=\frac{T_\cc^5}{cD^2} = \Big(\frac{c^2\xi^5}{D}\Big)^{\frac 13}
\label{eq:defLclowT}
\end{equation}
allowing to write the equation for the crossover in a adimensioned form as
\begin{equation}
   \frac{f_\cc}{f} = \Big(\frac{L}{L_\cc}\Big)^{\frac 43}
\end{equation}
{The relation \eqref{eq:defLclowT}} separates the flow and creep regimes depicted in Fig.~\ref{fig:phase-diagram} and Fig.~\ref{fig:2D_phase-diagram}.
For ${f_\cc}/{f} \ll ({L}/{L_\cc})^{4/3}$ the system is large enough for the global rescaling described in Sec.~\ref{ssec:scalingandthecreeplaw} to hold: the motion occurs by a succession of avalanches of size governed by $L_\opt(f)$. For  ${f_\cc}/{f} \gg ({L}/{L_\cc})^{4/3}$ this avalanche size is larger than the system size, and takes instead a size of the order of the system. The velocity then takes a linear form given by the low-$f$ expansion of~\eqref{eq:creep_1d_finiteL_sh} {at ${f\lesssim f_L}$}: the motion is a flow motion.

Note that, actually, ${L_\cc}$ is the low-temperature limit of the \emph{Larkin length} ${\mathcal{L}_\cc}$.
In our approach, we will use that ${\mathcal{L}_\cc}$ represents the lengthscale above which the Brownian rescaling is valid, as already hinted just before \eqref{eq:vftimeintegral}.
Historically, the {Larkin length} was introduced by Larkin and Ovchinnikov~\cite{larkin_pinning_1979} to describe the typical lengthscale of the geometrical fluctuations of driven vortices in type-II superconductors (see~\cite{blatter_vortices_1994} for a review).
It is also found to control the scale at which cusp singularity govern the renormalized disorder correlator in the FRG description of random manifold in dimension $4-\epsilon$ (see Refs.\cite{narayan_threshold_1993,chauve_creep_1998,chauve_creep_2000}).
We focus on its temperature dependence and its role in our one-dimensional problem in the next subsection.

It is appealing to identify $f_\cc$ with the critical depinning force $F_\cc$ of the depinning transition (Fig.~\ref{fig:interface}) but we have no argument in this favour, except if the creep energy barrier can be related to the unique energy barrier observed numerically and experimentally to describe the regime of force around the depinning~\cite{bolech_carlos_j._universal_2004,kolton_creep_2009,jeudy_universal_2016}. In our case, we can only affirm that $f_\cc$ represents the typical intrinsic energy density (along both transverse and longitudinal direction) of the interface which opposes to the constant drive.
We also emphasise that our results complement previous numerical and phenomenological works on the driven interface~\cite{kolton_creep_2009,ferrero_2013_ComptesRendusPhys14_641} and on their detailed scaling.

\subsection{Analysis of the scaling in temperature}
\label{ssec:tempescaling}

Last, we discuss how finite temperature affects the description of the phase diagram.
We first note that the Arrhenius low-temperature asymptotics regime $T \ll U_\cc \;(f_\cc/f)^{1/4}$ read from~\eqref{eq:creep_1d_TM} leaves room for a finite-temperature analysis of the dependency of the scales, thanks to the large factor $(f_\cc/f)^{1/4}$ (for $f\ll f_\cc)$.
In~\cite{agoritsas_static_2013} was derived the expression of a $T$-dependent Larkin length
\begin{equation}
  \mathcal L_\cc = \frac{T^5}{cD^2} g^{-5}
  \label{eq:TdepLarking}
\end{equation}
where $g$ is a dimensionless (``fudging'') parameter depending on $c,D,T,\xi$, denoted $f$ in~\cite{agoritsas_static_2013}, that distinguishes between a low-temperature ($T\ll T_\cc$) regime and a high-temperature regime ($T\gg T_\cc$).
The fudging parameter depends on the constants of the problem only through $T/T_\cc$.
At low $T$, one has $g\sim T/T_\cc$ and $\mathcal L_\cc $ becomes the length $L_\cc$ defined in~\eqref{eq:defLclowT} and used to adimensionalize $L$ in the phase diagram.
At high $T$, one has $g\to 1$ and the observables, such as $\mathcal L_\cc$, do not depend on $\xi$ anymore (one recovers the standard $\xi =0$ KPZ scalings).
{Physically, this means that at high temperature the typical energy scale is fixed by the temperature $T$, whereas in the $T\to 0$ limit where the thermal fluctuations are suppressed, it is the disorder which fixes the typical energy scale to ${T_\cc=(\xi c D)^{1/3}}$.}
The relation between $\tilde D$ and $g$ is
\begin{equation}
  \tilde D = g \frac{cD}{T}
  \label{eq:Dtildevsfudgingg}
\end{equation}
which is compatible with the asymptotic behaviour~\eqref{eq:lowhighTregimes}.
The crossover between the low- and high-temperature regimes is not known analytically, but several numerical, variational and scaling arguments were presented in~\cite{agoritsas_static_2013} supporting the fact that~$g$ interpolates smoothly between those two regimes.
{In particular, it was shown that the value ${g \in [0,1]}$ is directly related to the parameter governing of the full replica-symmetry-breaking of a Gaussian variational approach~\cite{agoritsas_temperature-induced_2010}.}

For the problem of interest, the length $\mathcal L_\cc$ is essential to delineate \emph{(i)} the regime $L\gg \mathcal L_\cc$ where the Brownian scaling that we have used for the scaling~\eqref{eq:rescaling_free-energy_FbarfV} of the disorder free-energy $\bar F$ is valid, from \emph{(ii)} the regime $L\ll \mathcal L_\cc$ where it is not.
At zero force, the line $L=\mathcal L_\cc$ which separates those two regimes corresponds to the equation
\begin{equation}
  \frac{T}{T_\cc} = \frac 1g \Big(\frac{L}{L_\cc}\Big)^{\frac 1 5} 
\label{eq:LLc_equiveq}
\end{equation}
%
This curve was obtained from an approximate equation obeyed by $g$ (derived in~\cite{agoritsas_static_2013}): depending on the approximation scheme, one has
\begin{equation}
  g^{\bar\gamma} = \Big(\frac{T}{T_\cc}\Big)^{\bar\gamma} (1-g) \qquad \text{with}\ {\bar\gamma}\in\{\tfrac 32,6\}
  \label{eq:fudging_eq}
\end{equation}
where we have omitted a dimensionless numerical prefactor that can be incorporated in the definition of $T_\cc$ without loss of generality.
The choice between the two possible exponents ${\bar\gamma}=6$ and ${\bar\gamma}=\frac 32$ affects the shape of the crossover but does not affect the dominant asymptotic behaviour of $g$ in the low- and high-temperature regimes. From~\eqref{eq:LLc_equiveq}, simple algebra allows to transform~\eqref{eq:fudging_eq} into an equation for the characteristic line $L=\mathcal L_\cc$ separating the Brownian and non-Brownian regimes in the plane $(\tfrac{L_\cc}{L},\frac{T}{T_\cc})$ of the phase diagram. One obtains:
\begin{equation}
  \frac{T}{T_\cc} = \frac{1-\big(\tfrac{L_\cc}{L}\big)^{\frac {\bar\gamma} 5}}{\big(\tfrac{L_\cc}{L}\big)^{\frac 15}}
 \label{eq:bluemanifold}
\end{equation}
The region $f>0$
can be studied in our approach for the regime of low forces where the quasistatic approach validating the effective model holds.
Thanks to the modified $f$-STS, the effective potential~\eqref{eq:STSwithf_forMFPT} is decomposed into a linear contribution in $f$ and a (distributionally) $f$-independent disorder, 
whose Brownian scaling at large $\tf$ is still governed by the condition $L\gg\mathcal L_\cc$ with the same $f$-independent $\mathcal L_\cc$.
In this regime of force, the demarcation between the Brownian and non-Brownian regime thus extends to the region $f>0$, as depicted in Fig.~\ref{fig:phase-diagram}.
For larger forces, closer to the regime of the depinning transition, it is known numerically that roughness exponent $\zeta$ differs from $\frac 23$ and one has to resort to other approaches~\cite{kolton_creep_2009,ferrero_2013_ComptesRendusPhys14_641} in order to describe scaling regimes that our effective model does not encompass.

In a similar way (see Fig.~\ref{fig:2D_phase-diagram}), $\mathcal L_\cc$ governs the regime of force in which $L_\opt(f)$ is large enough for the Brownian scaling of the free energy to hold.
Defining from $L_\opt(f_{\mathcal L}) = \mathcal L_\cc$, a Larkin force as
\begin{equation}
f_{\mathcal L} 
= \frac{\tilde D^7}{c^5 D^4}
= \frac{c^2 D^3}{(T/g)^7}
\quad\text{or equivalently}\quad
\frac{f_{\mathcal L}}{f_\cc} = \Big(g\frac{T_\cc}{T}\Big)^7
 \label{eq:eq-for-fL}
\end{equation}
one has that the Brownian rescaling holds only in the $f<f_{\mathcal L}$ regime.
The corresponding regimes are depicted at $T\to 0$ on Fig.~\ref{fig:2D_phase-diagram} and for all $T>0$ on Fig.~\ref{fig:phase-diagram}, where the green manifold defined by the condition $L_\opt(f) = \mathcal L_\cc$ is equivalently described by the equation
\begin{equation}
  \frac{T}{T_\cc} = \frac{1-\big(\tfrac{f}{f_\cc}\big)^{\frac {\bar\gamma} 7}}{\big(\tfrac{f}{f_\cc}\big)^{\frac 17}}
\label{eq:green-manifold}
\end{equation}
that follows from~\eqref{eq:fudging_eq}.

Finally, we examine the finite-$T$ dependence of $U_\cc$ implied by our analysis.
The low-temperature expression~\eqref{eq:UcfclowT} of $U_\cc$ and $f_\cc$ 
was deduced by direct identification between~\eqref{eq:finaltau1} and~\eqref{eq:creep_1d_TM} in the $T\ll T_\cc$ regime.
It can be extended at finite temperature: fixing $f_\cc$ to its temperature-independent expression~\eqref{eq:UcfclowT}, one obtains, by introducing the fudging parameter $g$, the following expressions
\begin{equation}
 U_\cc=\Big(g\frac{T_\cc}{T}\Big)^{\frac 34}T_\cc
 \:,\quad
 f_\cc
 =\Big(\frac{D^2}{c \xi^7}\Big)^{\frac 13}
\label{eq:UcfcfiniteT}
\end{equation}
which we expect to be valid in the  Arrhenius low-temperature regime (reading from~\eqref{eq:creep_1d_TM}: $T\ll U_\cc\;(f_\cc/f)^{1/4}$).
In the limit of $T\ll T_\cc$, ${g\sim T / T_\cc}$ we recover the relation ${U_\cc=T_\cc}$ of \eqref{eq:UcfclowT}.
The next order a small temperature expansion yields the following correction for $U_\cc$:
\begin{equation}
U_\cc= \Big(1-\frac{3}{4\bar\gamma}\frac{T}{T_\cc}\Big)T_\cc
\ + \ \mathcal{O}(T^{-2})
\end{equation}
We emphasise that the temperature dependence of $U_\cc$ depends on the definition of the characteristic force $f_\cc$, which we have chosen in \eqref{eq:UcfcfiniteT} to be temperature-independent.
This prediction still needs to be reconciled to the numerical results plotted in Fig.~3(b) of Ref.~\cite{kolton_creep_2005},
where the effective energy barrier has however been evaluated by identifying $f_\cc$ with the zero-temperature critical depinning force~$F_\cc$.
We also we note, importantly, that the Arrhenius low-temperature criterion
$T \ll U_\cc \;(f_\cc/f)^{1/4}$
for the instanton analysis to be valid, is, from~\eqref{eq:UcfcfiniteT}, equivalent to
\begin{equation}
  \frac{f}{f_\cc} \ll g^3  \Big(\frac{T_\cc}{T}\Big)^7
\quad\text{\emph{i.e.}, from~\eqref{eq:eq-for-fL}}:\quad
 f \ll g^{-4} f_{\mathcal L}
\end{equation}
which means that it is always satisfied when the mandatory condition $f \ll f_{\mathcal L}$ for the Brownian scaling to hold is satisfied.
We used here that the fudging parameter verifies $g\leq 1$ at all temperatures.
In other words, the low-temperature assumption for the low-noise analysis of the effective model to be valid is always satisfied as long as the condition $L_\opt(f)\gg \mathcal L_c$
, that guarantees the creep scaling, is verified.

\subsection{{Departure from the creep behaviour at intermediate forces}}
\label{ssec:fsize-intermediate}
Although the effective model is expected to be valid in the quasi-static regime only, we can study the first
correction that it brings to the pure creep regime at intermediate forces.
The scaling analysis shows that the effective Arrhenius barrier $\propto U_\cc (f/f_\cc)^{-1/4}$ actually decreases as the force increases (at $f>f_L$), 
implying that the backward contribution to the flow (the negative term of~\eqref{eq:creep_1d_finiteL-bis}) becomes more and more important as $f$ increases.
Such phenomenon is a consequence of the elasticity of the extended interface, that determines the scaling of the effective barrier, and does not occur for instance in the dynamics of a driven particle in a random potential (Sec.~\ref{sec:particle}).
The correction to the pure creep law is illustrated on Fig.~\ref{fig:barriers-modif-creep} (right): it induces a bending of the $\bar v(f)$ characteristics,
actually compatible with the experimental measurement of Ref.~\cite{lemerle_1998_PhysRevLett80_849}.

Numerical studies and experimental measurements on ferromagnetic domain walls \cite{kolton_creep_2009,bustingorry_kolton_2012_PhysRevB85_214416,gorchon_pinning-dependent_2014,jeudy_universal_2016}
have reported an intermediate affine (`TAFF'-like) regime succeeding to the creep regime at intermediate forces.
It would be interesting to compare such a regime to the correction to the creep regime due to the backward flow~\eqref{eq:creep_1d_finiteL_sh} (at $f>f_L$).
The experimental and numerical data of the above mentioned works are shown to be compatible with a law
\begin{equation}
  \bar v(f) \propto e^{\text{$-\frac{U_\cc}{T}\big[(f/f_\cc)^{- \frac 14}-1\big]$}}
  \label{eq:vf_vanishingbarrier}
\end{equation}
\emph{i.e.}~with an effective barrier $\propto \big[(f/f_\cc)^{- \frac 14}-1\big]$ that vanishes at $f=f_\cc$.
We argue that in the intermediate force regime the contribution of the backward flow are not negligible anymore: it would be interesting to determine whether the experimental evidence are indeed compatible with those contributions, by comparing measurements to~\eqref{eq:creep_1d_finiteL_sh} (at $f>f_L$) instead of~\eqref{eq:vf_vanishingbarrier}.
Note that both~\eqref{eq:creep_1d_finiteL_sh} and~\eqref{eq:vf_vanishingbarrier} describe an affine dependence of the velocity in $f$ in the intermediate regime force $f\lesssim f_c$, that is observed experimentally, but their origin is different.

\section{Discussion}
\label{sec:discussion}

Non-linear response laws correspond in our context to a glassy behaviour where metastable states are organised in a hierarchical manner~\cite{balents_bouchaud_mezard_1996_JPhysI6_1007}.
Phenomenological approaches can be problematic: as we have exposed in Sec.~\ref{ssec:naivescaling}, naive power counting arguments can lead to wrong results.
A remedy in such situations can be to derive effective models.
A tentative approach for the elastic line would be to consider a free-energy density of the driven line: for transverse fluctuations~$y$ at a scale~$L$, one would have to combine elastic, disorder and driving contributions, giving:
$F(y,L)=y^2L^{-2} + y^{\frac 1 2}L^{-1} - fy$ (setting to simplify all coefficients to one).
%
%
Choosing a power-law scaling $y=L^\zeta$ and equating the three terms yields the KPZ roughness exponents $\zeta_\KPZ=2/3$ together with the optimal driven length $L_\opt\sim f^{-3/4}$.
This description with \emph{one effective degree of freedom}~$y$ can however in no way constitute an effective model for the motion of the centre of mass $y$ of the interface, since the parabolic contribution $\propto y^2$ would confine it,
forbidding any long-time stationary driven regime with non-zero velocity. 

In contrast, the effective model we have defined in Sec.~\ref{ssec:effectivemodel} (see Fig.~\ref{fig:schemepoteff})
presents \emph{two effective degrees of freedom}: the extremities $\yi$ and $\yf$ of a segment of length $\tf$ of the interface, or alternatively its centre of mass $\bar y$ and the relative displacement $\deltay$.
Having this second degree of freedom $\deltay$, orthogonal to $\bar y$, now allows to include the parabolic contribution $\propto\deltay^2$ needed to encode the elasticity, without precluding the motion of the centre of mass (see Fig.~\ref{fig:schemepoteff}).
A first advantage of this effective model is that it was constructed explicitly from the original interface model, and that its contribution $\propto fy$ describing the effect of the force is derived through a generalisation of the STS (Appendix~\ref{sec:app_STS}).
This result is non-trivial in the sense that if linear response fails for the velocity, there is no reason \emph{a priori} that it would hold for the free energy, as is usually assumed in scaling arguments.
Besides,
the presence of the degree of freedom $\deltay$, orthogonal to the centre of mass $\bar y$, implies that the motion along direction $\bar y$ is not blocked by the highest barrier;
on the contrary, the motion can bypass those highest barriers by going through saddle points
(here one considers the low-temperature regime picture described in Sec.~\ref{ssec:MFPTeff}, where the motion is dominated by the instanton).
It corresponds for the original interface to the role of large elastic deformations, that become too costly might the interface be locally pinned by a strong fluctuation of the random potential.
Furthermore, the disordered contribution to the free-energy was also proven to be translationally invariant in distribution --~which, as we discussed, justifies the existence of a velocity at large time.
As a second advantage, the effective model is amenable to a large-$\tf$ scaling analysis that corresponds to the low-force asymptotics.
The power-counting result is thus justified by a saddle-point analysis at large $\tf$ of our effective model (subsection~\ref{ssec:scalingandthecreeplaw}),
extending the equilibrium saddle-point analysis of~\cite{agoritsas_static_2013}.

Another advantage of the effective model we put forward is that it allows a complete scaling analysis, even in the presence of short-range correlations at a scale~$\xi$ in the disorder.
Indeed, after rescaling, the free-energy becomes~\eqref{eq:scalingFfall} where the dependence in the parameters is gathered into a single prefactor, apart from disorder correlations which are rescaled to an effective lengthscale $\xi /a = \xi/(\frac{\tilde D}{cf})^{1/2}$.
In the small force limit $f\to 0$, this effective lengthscale goes to zero, and the disorder free-energy $\hat F$ in~\eqref{eq:scalingFfall} becomes the uncorrelated ($\xi=0$) one.
The dependence in the original correlation length $\xi$ is absorbed in the common prefactor to all terms of~\eqref{eq:scalingFfall}, through the constant $\tilde D$~\cite{agoritsas_static_2013,agoritsas_temperature-induced_2010,agoritsas_disordered_2012}.
This global scaling properties support that, as often informally stated, all barriers of the creep problem scale in the same way proportionally to $f^{-1/4}$.
It would be interesting to relate such picture to the one of a single energy barrier recently shown in~\cite{jeudy_universal_2016} to correctly describe the velocity-force dependence  beyond the creep regime, in experimental and numerical results.

We can make the connection between our results and a special regime of the KPZ fluctuations, using that at~$\xi=0$ results more precise than the Brownian scaling~\eqref{eq:rescaling_free-energy_FbarfV} at $\tf\to\infty$ are available. Indeed, at $\xi=0$ the disorder free-energy scales as follows (\cite{amir_probability_2011,calabrese_free-energy_2010,dotsenko_bethe_2010}, see~\cite{corwin_kardarparisizhang_2012} for a review):
\begin{equation}
  \bar F_V(\tf,\yf|0,\yi)  
  \stackrel{\text{(d)}}{=} 
  \Big(\frac{\tilde D^2}{c}\tf\Big)^{\frac 13}
  \,\mathcal A_2\Big((\yf-\yi)/[(\tilde D/c^2)^{1/3} \tf^{2/3}]\Big)
 \label{eq:free-energy_FbarfV_Airy}
\end{equation}
where $\mathcal A_2$ is the Airy$_2$ process and $\tilde D = \frac{cD}{T}$.
We thus have that upon the rescaling~\eqref{eq:ytab} with the choice~\eqref{eq:rescaclassical}
\begin{equation}
  \bar F_V(\tf,\yf|0,\yi) 
  \stackrel{\text{(d)}}{=} 
  \Big(\frac{\tilde D^2 b}{c}\htf\Big)^{\frac 13}
  \, \mathcal A_2\big([\hyf-\hyi]/\htf^{2/3}\big)
 \label{eq:resc_free-energy_FbarfV_Airy}
\end{equation}
Finally, upon the same rescaling~\eqref{eq:rescbf} as in the Brownian case for the scale of time~$b$ as a function  of the force $f$, we obtain that~\eqref{eq:scalingFfall} becomes
\begin{equation}
  F^f_V(\tf,\yf|0,\yi) 
=
  f^{-\frac 14}
  \Big(\frac{\tilde D^3 }{c}\Big)^{\frac 14}
\Big[
  \frac{(\hyf-\hyi)^2}{2\hat\tf}  + \htf^{\frac 13} \; \mathcal A_2\big([\hyf-\hyi]/\htf^{2/3}\big)
-
   \frac {\hat  \tf}2  (\hyf+\hyi) 
\Big]
\label{eq:scalingFfall_Airy2}
\end{equation}
a form where the dependence in the physical parameters has been absorbed in a unique prefactor.
The rest of the analysis presented in Sec.~\ref{ssec:scalingandthecreeplaw} remains formally valid: the saddle-point asymptotics in the low-force limit $f\to 0$ remains dominated by optimal values for $\hyi$, $\hyf$, $\htf$ which do not depend on the parameters, and the generic form of the creep law~\eqref{eq:creep_1d_TM} is also recovered.
The issue is that the scaling in distribution~\eqref{eq:free-energy_FbarfV_Airy} has been shown to be valid only in the strict $\xi=0$ case, which corresponds to the regime $T\gg T_\cc$ of our settings.
It has been however evidenced numerically in Ref.~\cite{agoritsas_finite-temperature_2012} that~\eqref{eq:free-energy_FbarfV_Airy} remains valid at finite $\xi$ for $|\yf-\yi|\gtrsim \xi$, provided $\tilde D$ is replaced by its finite-$\xi$ counterpart as in~(\ref{eq:rescaling_free-energy_FbarfV}-\ref{eq:lowhighTregimes}), in the equilibrium $f=0$ case.
Our analysis thus provides motivation to study the extension of~\eqref{eq:free-energy_FbarfV_Airy} to the finite-$\xi$ case in further details, since, if valid, it remains compatible with the scaling of the creep law at $f\neq 0$.

Besides, the results we have presented are complementary to previous studies of elastic interfaces in random media with long-range elasticity~\cite{tanguy_weak_2004,patinet_quantitative_2013}.
We can define the ratio $\Xi_0=D^{1/2}/c$, which allows to rewrite the finite-size coordinate of the phase diagram (Fig.~\ref{fig:phase-diagram}) as $L_\cc/L = (\xi/\Xi_0)^{2/3} \xi/L$.
(Note that $\Xi_0$ is not in general a length, unless one chooses a system of units in which the transverse and longitudinal directions have same dimensions --~which is of course the case for the interface, but not in the directed-polymer or KPZ description).
Then, \emph{at fixed $\xi$ and fixed $L$}, the critical region at small $L_\cc/L$, where the creep law holds, corresponds to the large $\Xi_0$ asymptotics (\emph{i.e.}~`strong pinning': as seen from the expression of $\Xi_0$, disorder dominates elasticity), while the asymptotics at small $L_\cc/L$, outside of the creep regime, corresponds to `weak pinning'~(see \cite{larkin_pinning_1979} for strong and weak pinning in type-II superconductors, and the review~\cite{giamarchi_statics_1998}).

Last, we emphasise that our analysis of the stationary velocity of the effective model in its non-equilibrium steady state was made possible through a mean first passage time (MFPT) argument. As we discussed, it allowed us to obtain $\bar v(f)$ within boundary conditions that would induce an \emph{equilibrium} steady-state, but through the determination of a non-stationary MFPT.
Although powerful (since it transforms a non-reversible problem into a reversible one), such reasoning will fail at large force, and another approach should be developed to understand this regime.
We also note that other types of effective models have been used in the context of interfaces with long-range elasticity~\cite{patinet_quantitative_2013} and it would be interesting to establish connections to the results.

\section{Conclusion}
\label{sec:conclusion}

\subsection{Summary}

{
The derivation of the velocity-force dependence relies on the combination of three distinct limits:
\textit{(i)}~low temperature $T$, \textit{(ii)}~large system size $L$, and \textit{(iii)}~small forces $f$.
The understanding of the characteristic scales defining those limits in our construction has allowed us to grasp the validity range of the creep regime and to identify how it is modified at intermediate forces.
}

{
The low-temperature assumption allows for the Arrhenius MFPT expression \eqref{eq:restau1_non-ave} based on the instanton description.
It is valid in the limit where the temperature $T$ is very small in comparison to the effective barrier $U_\cc \;(f_\cc/f)^{1/4}$.
Because of large prefactor $\propto(f_\cc/f)^{1/4}$ at $f\ll f_\cc$, the domain of validity of the instanton description extends much beyond the naive regime $T\ll U_\cc$ (which reads $T\ll T_\cc$ at low temperature), allowing in particular to study  in a well defined way the dependence of the scale $U_\cc$ as $T$ varies.
Increasing the temperature modifies the geometrical fluctuations of the polymer --~both their amplitude (related to $\tilde{D}$) and their characteristic lengthscales (such as $\mathcal{L}_\cc$)~--
with a temperature dependence parametrised by the fudging parameter $g$ as presented in Sec.~\ref{ssec:tempescaling}. 
This affects the scaling of the Larkin length $\mathcal L_\cc$: at low temperature we have ${\mathcal{L}_\cc \approx L_\cc \sim \frac{T_\cc^5}{cD^2}}$, whereas in the opposite limit ${T \gg T_\cc}$ we have ${\mathcal{L}_\cc \sim \frac{T^5}{cD^2}}$ \cite{agoritsas_static_2013}.
This has allowed us to determine, in the $(L_\cc/L,T/T_\cc)$ plane, the region where the Brownian scaling of the free energy holds.
To characterise higher temperatures ($T\sim U_\cc \;(f_\cc/f)^{1/4}$) where the Arrhenius description breaks down, one would need to take into account the contributions of the fluctuations around the instanton and of the other transition paths (for instance through Morse theory~\cite{milnor_morse_1973,tanase-nicola_metastable_2004}).
}

{
Within the validity range of the Arrhenius and instanton description, the system size should be sufficiently large (${L \gg \mathcal{L}_\cc}$) so that the MFPT expression \eqref{eq:vftimeintegral} is dominated by the  Brownian scaling of the disorder free energy and the KPZ scaling of the geometrical fluctuations (${y(\tf)^2 \sim \tf^{4/3}}$).
For systems smaller than the Larkin length (${L < \mathcal{L}_\cc}$), the Brownian rescaling of the free-energy is not valid any more. This implies that our study of the creep regime breaks down and that the sub-Larkin scalings of the geometrical fluctuations ${y(\tf)^2 \sim \tf^{2 \zeta}}$ with ${\zeta \neq \tfrac 23}$ will affect the free-energy rescalings. In this regime, the value of the roughness exponent is actually unknown from an analytical point of view, but it has been evaluated to be larger than 1 \cite{agoritsas_static_2013,agoritsas_staticnum_2013}.
}

{
The first implication of the small force assumption is that the ${f \to 0}$ asymptotics corresponds to the  ${\vmoy(f) \to 0}$ asymptotic, allowing for the quasistatic approximation.
We have explicitly implemented the latter in our effective model \eqref{eq:effdynf} by using the \emph{static} free energy (see Sec.~\ref{ssec:effectivemodel}).
Although the STS for the static free energy at finite force remains valid at an arbitrary large force (see Appendix~\ref{ssec:STS_nonzero-force}), its use for the effective model is self-consistent only in the ${f \to 0}$ asymptotics.
The second implication of considering this asymptotics is that, after performing the Brownian rescaling~\eqref{eq:rescaling_free-energy_FbarfV} allowed by ${L > \mathcal{L}_\cc}$, it is possible to perform a saddle-point argument for the MFPT \eqref{eq:finaltau1} and to infer from it the steady-state velocity \eqref{eq:creep_1d_TM}.
The validity range of the creep regime is thus restricted to ${\mathcal{L}_\cc < L_{\opt}(f) < L}$ with ${L_{\opt}(f) \sim  f^{-3/4}}$.
Decreasing the force, one eventually reaches ${L_\opt(f_L) \equiv L}$ and observes the finite-size crossover discussed in Sec.~\ref{ssec:fsize}.
Increasing the force, ${L_\opt(f)}$ decreases and when one reaches ${L_\opt(f_{\text{max}}) \equiv \mathcal{L}_\cc}$ the Brownian scaling~\eqref{eq:rescaling_free-energy_FbarfV} cannot be used anymore to rescale the free energy in the MFPT expression \eqref{eq:vftimeintegral}. At low temperature we have ${f_{\text{max}}=f_\cc}$, confining the creep regime at most to forces ${f \in \left[f_L , f_\cc \right]}$, whose scalings are known only to numerical prefactors that we cannot access in our approach.
}
The third implication is that in the small force regime, since the effective barrier $U_\cc\;(f_\cc/f)^{1/4}$ increases as $f\to 0$, one can neglect the backward flow compared to the forward flow, as discussed in Sec.~\ref{ssec:fsize-intermediate}: the negative contribution to~\eqref{eq:creep_1d_finiteL-bis} is negligible compared to the positive one, which yields the pure creep law~\eqref{eq:creep_1d_TM}.

\subsection{Perspective}
We have {proposed} and studied a two-degrees-of-freedom effective model describing the motion of a driven 1D elastic line in a disordered random medium.
Through a mean-first-passage-time study and a saddle-point argument at small driving force, we provided a detailed derivation of the creep law {(see \eqref{eq:finaltau1}, \eqref{eq:creep_1d_TM} and \eqref{eq:UcfcfiniteT})}
, and we used our proposed analysis to understand the crossover from the creep law to the linear response regime that one expects at very small forces for finite systems {(see \eqref{eq:creep_1d_finiteL_sh})}.
We established the phase diagram which describes this crossover together with the critical region where the creep law is valid.

Extensions of the approach we have described could be interesting to understand the dynamics and jump statistics of driven vortices in superconductors presenting dislocation planes (experiments of~\cite{shapira_disorder-induced_2015}), in relation to the recent theoretical work of Ref.~\cite{aragon_avalanches_2016}.
Other experiments (this time for interfaces in magnetic materials~\cite{jeudy_universal_2016,gorchon_pinning-dependent_2014}) are compatible with a thermally-assisted flux flow (TAFF) at \emph{intermediate driving force} (below the depinning force, but beyond the creep regime). It would be worth trying to extend our analysis to this regime of force, but it would require to incorporate a roughness different from $\zeta_\KPZ=\frac 23$.
The existence of three flow regimes in the velocity-force characteristics (at finite size ${L \gg L_\opt}$ described in this article; the intermediate force TAFF; and the large-force ``fast flow'' regime at $f\gg f_\cc$) also rises the question of whether and, if so, how those regimes are connected.
Mathematical aspects are also worth investigating: an effective model with one degree of freedom, quadratic elasticity and Brownian disorder corresponds to the Brox diffusion, see Ref.~\cite{brox_one-dimensional_1986}. The extension to two degrees of freedom and the inclusion of a driving field might allow to reach temperatures beyond the zero-noise limit that we had to restrict to.

Our analysis relies on the decomposition of the free-energy into contributions which rescale in a simple manner; in one dimension, the existence of this decomposition is related to the Airy process which describes the large $\tf$ behaviour of the directed-polymer free-energy scaling.
In two dimensions, numerical simulations provide evidences supporting the existence of a corresponding universal process~\cite{halpin-healy_2+1-dimensional_2012,halpin-healy_extremal_2013}, the scaling of which supposedly controlling the low-force regime.
More complex descriptions are sometimes required to understand the dynamics of magnetic domain walls, including an internal phase which plays the role of a hidden internal degree of freedom~\cite{le_doussal_depinning_2008,lecomte_depinning_2009,stewart_e_barnes_and_jean-pierre_eckmann_and_thierry_giamarchi_and_vivien_lecomte_noise_2012}.
The role of disorder in the dynamics of such systems is poorly understood, either in the field-driven or the current-driven cases, excepted in zero-dimensional toy models~; a generalisation of the approach we have presented could be instructive.

\begin{acknowledgements}
%
%
{We thank Thierry Giamarchi for fruitful discussions, especially in the early stages of this work.}
E.A.~acknowledges financial support by a Fellowship for Prospective Researchers Grant No P2GEP2-15586 from the Swiss National Science Foundation.
V.L.~thanks the Montbrun-les-Bains Centre for Theoretical Physics for its warm hospitality.
E.A.~and V.L.~ackowledge support by the National Science Foundation under Grant No. NSF PHY11-25915 during a stay at KITP, UCSB where part of this research was performed.
R.G.G.~acknowledges financial support from Labex LaSIPS (No. ANR-10-LABX-0040-LaSIPS), managed by the French National Research Agency under the “Investissements d’avenir” program (No.
ANR-11-IDEX-0003-02).
\end{acknowledgements}


\appendix
\renewcommand{\theequation}{\Alph{section}.\arabic{equation}}

\section{Statistical Tilt Symmetry at $f\neq 0$}
\label{sec:app_STS}

We derive in this appendix a form of the Statistical Tilt Symmetry (STS), which allows to decompose the point-to-point free energy into the sum of a deterministic contribution and of disorder-dependent one, which, in distribution, is invariant by translation along direction~$y$.
We first recall the known result at zero force before deriving a novel STS at $f\neq 0$.

\subsection{A reminder: the zero force STS ($f=0$)}
\label{ssec:STS_zero-force}

In the zero force situation, the linear change of coordinates $y(t)= \tilde y(t) + \yi+ \frac{\yf-\yi}{\tf} t$
allows to relate the weight of trajectories starting in $(0,\yi)$ and arriving in $(\tf,\yf)$ (see Fig.~\ref{fig:interface}) to those starting in $(0,0)$ and arriving in $(\tf,0)$
in a translated disorder.
One directly reads from~\eqref{eq:defWVf} (with $f=0$) that
\begin{equation}
   W_V(\tf,\yf|0,\yi) = \ee^{-\frac cT  \frac{(\yf-\yi)^2}{2 \tf}} W_{\T_{\yfi}^{\tf} V}(\tf,0|0,0)
\end{equation}
where the translated disorder $\T_{\yfi}^{\tf} V$ is defined as
\begin{equation}
 \T_{\yfi}^{\tf} V(t,\tilde y)\equiv V\big(t,\tilde  y+\yi+\tfrac{\yf-\yi}{\tf}t\big) 
\end{equation}
This remark enables to decompose the free energy $F_V(\tf,\yf|0,\yi)=-T\log W_V(\tf,\yf|0,\yi)$ as
\begin{equation}
  F_V(\tf,\yf|0,\yi) = F_{V\equiv 0}(\tf,\yf-\yi) + \bar F_V(\tf,\yf|0,\yi)
  \label{eq:STSf0}
\end{equation}
where  
\begin{align}
  F_{V\equiv 0}(\tf,\yf-\yi) &= F_{\text{t\hspace*{-.16mm}h}}(\tf,\yf-\yi) + \frac T2  \log \frac{2\pi Tt}{c}
\\
  F_{\text{t\hspace*{-.16mm}h}}(\tf,\yf-\yi) &=  c \frac{(\yf-\yi)^2}{2 \tf}
\end{align}
Hence, $\bar F_V(\tf,\yf|0,\yi)$ is invariant by translation in distribution, as we indeed observe that it is depending on $\yi$ and $\yf$ only through $\T_{\yfi}^{\tf}V$.
\begin{equation}
\bar F_V(\tf,\yf|0,\yi)
=
-T \log W_{\T_{\yfi}^{\tf} V}(\tf,0|0,0)
\end{equation}

\subsection{The non-zero force STS ($f\neq 0$)}
\label{ssec:STS_nonzero-force}

In the non-zero force situation, the key observation is that there exists a $f$-dependent transformation of the directed polymer trajectories
\begin{equation}
y(t)= \tilde y(t) + \yi+ \frac{\yf-\yi}{\tf} t + \frac{f}{2c}t(\tf-t)
\label{eq:tilt_fSTS}
\end{equation}
which, remarkably, implies a $f$-STS generalising~\eqref{eq:STSf0}:
\begin{equation}
   W^f_V(\tf,\yf|0,\yi) = 
\exp \Big\{
-\frac 1T
\Big[
c \frac{(\yf-\yi)^2}{2 \tf}-\frac {f}2 \tf (\yf+\yi) -\frac{f^2}{24 c}\tf^3
\Big]
\Big\} 
W_{\T_{\yfi}^{\tf,f} V}(\tf,0|0,0)
    \label{eq:STSwithf_TV}
\end{equation}
%
The result is obtained by coming back to the path-integral definition of the $f$-dependent point-to-point partition function~\eqref{eq:defWVf}, and taking care of the boundary terms.
The key observation in the computation is to recognise the following total derivative
\begin{equation}
\frac{ (2 c (\yf-\yi)+f \tf (\tf-2 t))}{2 \tf}\partial_t \tilde y(t)-f \tilde y(t) = 
\partial_t\Big[
\frac{f \tf ( \tf - 2 t) + 2 c (\yf - \yi) }{2 \tf} \tilde y(t)
\Big]
\end{equation}
for the terms linear in $\tilde y(t)$ after the change of variable~\eqref{eq:STSwithf_TV}.
The translated disorder $\T_{\yfi}^{\tf,f} \Vf$ is defined as
\begin{equation}
 \T_{\yfi}^{\tf,f} V(t,\tilde y)\equiv V\big(t,\tilde  y+\yi+\tfrac{\yf-\yi}{\tf}t+ \tfrac{f}{2c}t(\tf-t)\big) 
 \label{eq:transfoVdepf}
\end{equation}
Note importantly that if the translated disorder~\eqref{eq:transfoVdepf} depends on the force through a $f$-dependent change of coordinate, 
the partition function on the right hand side of~\eqref{eq:STSwithf_TV} is however the one \emph{at zero force}.
This enables to decompose the free energy $F_V^f(t,y)$ as
\begin{equation}
  F_V^f(\tf,\yf|0,\yi) = F_{\text{t\hspace*{-.16mm}h}}(\tf,\yf-\yi) -\frac {f\tf}2 (\yf+\yi) -\frac{f^2}{24 c}\tf^3
 + \bar F^f_V(\tf,\yf|0,\yi)
+
 \frac T2  \log \frac{2\pi Tt}{c}
  \label{eq:STSwithf}
\end{equation}
where 
\begin{equation}
\bar F^f_V(\tf,\yf|0,\yi)
=
-T \log W_{\T_{\yfi}^{\tf,f} V}(\tf,0|0,0)
  \label{eq:FfVWVf0}
\end{equation}
For uncorrelated disorder ($\xi=0$) this implies in particular that at large $\tf$ the distribution of $\bar F^f_V(\tf,\yf|0,\yi)$ adopts the Airy$_2$ scaling~\cite{corwin_kardarparisizhang_2012} and goes to the distribution of a Brownian motion of coordinate $\yf-\yi$ at infinite $\tf$ (this corresponds to the steady state of the KPZ equation~\cite{huse_huse_1985}).
For correlated disorder ($\xi>0$), the picture is slightly changed at large $\tf$: the steady state is not distributed as a Brownian anymore but scales similarly (as long as $\yf-\yi\gg \xi$)  with a different $\xi$-dependent amplitude~$\tilde D$ \cite{agoritsas_static_2013,agoritsas_staticnum_2013} (see Eq.~\eqref{eq:rescaling_free-energy_FbarfV} in the main text).

\section{Effective temperature $\tilde T$ and friction $\tilde \gamma$ in the absence of disorder}
\label{app:effectiveTgamma}

In this Appendix, we determine the relation between the original friction $\gamma$ and temperature~$T$ of the interface dynamics~\eqref{eq:Langevin_yttau_forcef} and the ones $\tilde\gamma$ and $\tilde T$ of the effective model~(\ref{eq:effdyni}-\ref{eq:effdynf}), in the absence of disorder ($V\equiv 0$).
Starting by the original dynamics, one averages~\eqref{eq:Langevin_yttau_forcef} over thermal fluctuations and taking the long-time limit, one gets
\begin{equation}
  \vmoy(f)\big|_{V\equiv 0} = \frac{f}{\gamma}
  \label{eq:vfV0}
\end{equation}
Changing to the reference frame of the centre of mass, one recognises that $y(t,\tau)-\frac{f}{\gamma}\tau$ obeys the Edwards-Wilkinson equation, whose steady state is Gaussian, implying that
\begin{equation}
  \big\langle[y(\tf)-y(0)]^2\big\rangle\big|_{V\equiv 0} = \frac T c\,\tf
  \label{eq:variancedeltayoriginal}
\end{equation}

On the other hand, the effective equations~(\ref{eq:effdyni}-\ref{eq:effdynf}) are written as follows in the absence of disorder, as seen from the expression~\eqref{eq:STSwithf} of the tilted free energy:
\begin{align}
\tilde\gamma \partial_\tau \yi(\tau)
& \ = \
+c\frac{\yf-\yi}{\tf}+\frac 12 f\,\tf 
+
\sqrt{2\tilde\gamma\,\tilde T}\,\tilde\eta_\ii(\tau)
\label{eq:effdyniV0}
\\
\tilde\gamma \partial_\tau \yf(\tau)
& \ = \
-c\frac{\yf-\yi}{\tf}+\frac 12 f\,\tf 
+
\sqrt{2\tilde\gamma\,\tilde T}\,\tilde\eta_\ff(\tau)
\label{eq:effdynfV0}
\end{align}
Summing these equations, averaging over thermal noise and imposing that both $\yi(\tau)$ and $\yi(\tau)$ move at an average velocity $\vmoy(f)|_{V\equiv 0}$ at large times, one 
gets~$2\tilde\gamma\,\vmoy(f)|_{V\equiv 0} = f\tf$; hence, comparing to~\eqref{eq:vfV0}, one obtains
\begin{equation}
  \tilde\gamma \ = \ \frac 12 \tf\,\gamma
  \label{eq:relationgammagammatilde}
\end{equation}
Note that to get this result, we relaxed the conditions (walls) that would enforce the model to reach a zero-velocity equilibrium steady state at large times. However, a MFPT analysis in such conditions would also yield the result~\eqref{eq:relationgammagammatilde} for the velocity in a finite window of the system far from the walls.
Subtracting now~\eqref{eq:effdyniV0} and~\eqref{eq:effdynfV0} one obtains a closed equation for $\deltay(\tau)=\yf(\tau)-\yi(\tau)$ in the absence of disorder:
\begin{equation}
  \tilde\gamma\partial_\tau\deltay(\tau)
=
  -2\frac{c}{\tf}\deltay(\tau) + \sqrt{4\tilde\gamma\tilde T}\,\tilde\eta_1
\end{equation}
where $\tilde\eta_1$ is a white noise of unit variance.
The force term of this Langevin equation derives from the energy $\frac{c}{\tf}(\deltay)^2$ and the noise term has temperature $2\tilde T$.
This shows that the steady-state distribution of $\deltay$ is Gaussian $\propto \exp[-\frac{c}{2\tilde T \tf}(\deltay)^2]$. This implies in turn that
$
\langle(\deltay)^2\rangle|_{V\equiv 0}=\langle[\yf-\yi]^2\rangle|_{V\equiv 0}=\frac{\tilde T}{c}\tf
$.
Finally, comparing to~\eqref{eq:variancedeltayoriginal} one obtains that the temperatures of the original and of the effective models are equal:
\begin{equation}
  \tilde T = T
\label{eq:TtildeequalsT}
\end{equation}
To summarise, in the absence of disorder, the effective model with friction~\eqref{eq:relationgammagammatilde} and temperature~\eqref{eq:TtildeequalsT} presents the same velocity and the same Gaussian end-point distribution as the original dynamics. We assume that this correspondence between original and effective parameters also holds in the presence of disorder.


\addcontentsline{toc}{section}{References}
\bibliographystyle{plain_url}       
\bibliography{DP-and-creep_article}   

\end{document}